\documentclass[acmtog,nonacm]{acmart}

\usepackage{booktabs} 
\usepackage{multirow}
\usepackage{amsmath}

\usepackage{graphicx}

\usepackage{caption}
\usepackage{subfigure}
\usepackage{algorithmic}
\usepackage{xspace}
\usepackage[table]{xcolor}
\usepackage[ruled]{algorithm2e}

\SetAlFnt{\small}
\SetAlCapFnt{\small}
\SetAlCapNameFnt{\small}
\SetAlCapHSkip{0pt}

\usepackage{cleveref}
\usepackage{dsfont}

\usepackage{amsmath,amsfonts,bm}

\def\eqref#1{equation~\ref{#1}}

\def\1{\bm{1}}

\def\rva{{\mathbf{a}}}

\def\rvd{{\mathbf{d}}}

\def\rvg{{\mathbf{g}}}
\def\rvh{{\mathbf{h}}}

\def\rvp{{\mathbf{p}}}
\def\rvq{{\mathbf{q}}}

\def\rvs{{\mathbf{s}}}

\def\rvx{{\mathbf{x}}}

\def\rvz{{\mathbf{z}}}

\DeclareMathAlphabet{\mathsfit}{\encodingdefault}{\sfdefault}{m}{sl}
\SetMathAlphabet{\mathsfit}{bold}{\encodingdefault}{\sfdefault}{bx}{n}

\def\sK{{\mathbb{K}}}

\def\sR{{\mathbb{R}}}

\usepackage[normalem]{ulem}
\usepackage{enumitem}

\definecolor{codegreen}{rgb}{0,0.6,0}
\definecolor{codegray}{rgb}{0.5,0.5,0.5}
\definecolor{codepurple}{rgb}{0.58,0,0.82}
\definecolor{backcolour}{rgb}{0.95,0.95,0.92}

\acmJournal{TOG}

\begin{document}

\title{SMP: Reusable Score-Matching Motion Priors for Physics-Based Character Control}

\author{Yuxuan Mu}
\authornote{Joint First Authors.}
\email{yma101@sfu.ca}
\affiliation{%
  \institution{Simon Fraser University}
  \country{Canada}
}

\author{Ziyu Zhang}
\authornotemark[1]
\email{zza333@sfu.ca}
\affiliation{%
  \institution{Simon Fraser University}
  \country{Canada}
}

\author{Yi Shi}
\authornotemark[1]
\email{ysa273@sfu.ca}
\affiliation{%
  \institution{Simon Fraser University}
  \country{Canada}
}

\author{Dun Yang}
\email{dya66@sfu.ca}
\affiliation{%
  \institution{Simon Fraser University}
  \country{Canada}
}

\author{Minami Matsumoto}
\email{Minami.X.Matsumoto@sony.com}
\affiliation{%
  \institution{Sony Interactive Entertainment}
  \country{Japan}
}

\author{Kotaro Imamura}
\email{Kotaro.Imamura@sony.com}
\affiliation{%
  \institution{Sony Interactive Entertainment}
  \country{Japan}
}

\author{Guy Tevet}
\email{guy.tvt@gmail.com}
\affiliation{%
  \institution{Stanford University}
  \country{USA}
}

\author{Chuan Guo}
\email{guochuan5513@gmail.com}
\affiliation{%
  \institution{Snap Inc.}
  \country{USA}
}

\author{Michael Taylor}
\email{Mike.Taylor@sony.com}
\affiliation{%
  \institution{Sony Interactive Entertainment}
  \country{USA}
}

\author{Chang Shu}
\email{Chang.Shu@nrc-cnrc.gc.ca}
\affiliation{%
  \institution{National Research Council Canada}
  \country{Canada}
}

\author{Pengcheng Xi}
\email{Pengcheng.Xi@nrc-cnrc.gc.ca}
\affiliation{%
  \institution{National Research Council Canada}
  \country{Canada}
}

\author{Xue Bin Peng}
\email{xbpeng@sfu.ca}
\affiliation{%
  \institution{Simon Fraser University}
  \country{Canada}
}
\affiliation{%
  \institution{NVIDIA}
  \country{Canada}
}

\begin{teaserfigure}
  \centering
  \includegraphics[width=\textwidth]{figures/demos/teaser.png}
  \caption{
  Our framework constructs reusable and modular motion priors for training motion controllers. A general motion prior can be trained on a large dataset spanning 100 distinct styles. Once trained, the prior can be repurposed into 100 style-specific priors without requiring access to the original dataset or further model updates. These style priors serve as stationary reward models and can be reused to train control policies for diverse tasks with natural stylistic behaviors.
  }
  \label{fig:teaser}
  \vspace{2mm}
\end{teaserfigure}

\begin{abstract}
    Data-driven motion priors that can guide agents toward producing naturalistic behaviors play a pivotal role in creating life-like virtual characters. Adversarial imitation learning has been a highly effective method for learning motion priors from reference motion data. However, adversarial priors, with few exceptions, need to be retrained for each new controller, thereby limiting their reusability and necessitating the retention of the reference motion data when applied to downstream tasks. In this work, we present Score-Matching Motion Priors (SMP), which leverages pre-trained motion diffusion models and score distillation sampling (SDS) to create reusable task-agnostic motion priors. SMPs can be pre-trained on a motion dataset, independent of any control policy or task. Once trained, SMPs can be kept frozen and reused as general-purpose reward functions to train new policies to produce naturalistic behaviors for downstream tasks. We show that a general motion prior trained on large-scale datasets can be repurposed into a variety of style-specific priors. Furthermore, SMP can compose different styles to synthesize new styles not present in the original dataset. Our method can create reusable and modular motion priors that produce high-quality motions comparable to state-of-the-art adversarial imitation learning methods. In our experiments, we demonstrate the effectiveness of SMP across a diverse suite of control tasks with physically simulated humanoid characters. Video available at \href{https://youtu.be/jBA2tWk6vzU}{youtu.be/jBA2tWk6vzU}.
\end{abstract}
%
%


%
%

\keywords{character animation, score distillation sampling, diffusion model, reinforcement learning}

\maketitle

\section{Introduction}
\label{sec:intro}

Creating virtual characters that move with natural and life-like behaviors is fundamental to immersive digital experiences in animation, films, games, and virtual reality (VR) applications. While motion capture and procedural animation techniques can produce high-quality movements, the challenge lies in developing controller that enable physically simulated characters to exhibit similarly natural behaviors dynamically, in response to diverse tasks and environmental contexts. Traditional physics-based controllers trained without reference motion data often result in visually unnatural movements, lacking the subtle qualities of life-like motions~\citep{yin2007simbicon, hodgins1995animating, 2010-TOG-gbwc}. 
Tracking-based methods improve realism by following reference motion clips, but they typically require controllers to rigidly mimic target motions frame-by-frame~\citep{peng2018deepmimic, liu2017learning, won2017train}, limiting their flexibility to deviate from the reference and adapt behaviors to perform new tasks.
Alternatively, distribution-matching methods, such as adversarial imitation learning~\citep{peng2021amp,ho2016generative}, provide a more versatile approach for imitating motion data. Adversarial methods can learn flexible motion priors from motion dataset, which can act as task-agnostic measures of motion naturalness, allowing policies to produce behaviors that resemble natural motions across different tasks. However, adversarial methods typically require a prior (\emph{i.e.} discriminator) to be trained jointly with a given policy, which limits the reusability and modularity of the learned priors. For each new policy, the prior must be continuously trained with data from the policy and data from the original reference dataset. As a result, the original dataset must be retained for perpetuity. 

We contend that for many practical control applications, an ideal motion prior should be \textbf{modular} and \textbf{reusable}:
\begin{itemize}[leftmargin=2em]
    \item \textbf{Modular}: The prior should function as an independent motion-quality objective that can guide policy training without requiring access to the original reference dataset.
    \item \textbf{Reusable}: Once constructed, the same prior should be applicable across diverse tasks and multiple policies without requiring further training.
\end{itemize}
 To develop motion priors that satisfy these criteria, we propose Score-Matching Motion Priors (SMP): a method for constructing reusable, modular motion priors that can be utilized to train control policies to produce naturalistic behaviors across diverse tasks. Given an unstructured motion dataset, we first train a motion diffusion model to capture the underlying data distribution independent of any task or control policy. Once trained, the diffusion model is kept frozen and repurposed as a prior via score distillation sampling (SDS)~\citep{poole2022dreamfusion}, providing a robust similarity measure between simulated motions and the behaviors in the reference dataset. By reusing the learned SMP as a general style reward, our system enables a prior to be used to train new policies to perform a diverse suite of tasks with natural life-like behaviors.

The central contribution of this work is a method for constructing \emph{modular} and \emph{reusable} motion priors for physics-based character animation by leveraging diffusion models as reward models in reinforcement learning. Our framework produces high-quality motions comparable to state-of-the-art adversarial imitation learning methods, while substantially improving modularity and reusability. 
SMP enables our system to completely discard the reference dataset during policy training.
We show that a score-matching motion prior can be constructed from a task-agnostic diffusion model pretrained on large-scale motion datasets. This general-purpose motion prior can then be repurposed into style-specific priors through prompting and guidance. Furthermore, we show that priors for different styles can be composed to create new behavioral styles that are not present in the original dataset. Code is available in \href{https://github.com/xbpeng/MimicKit}{MimicKit}.

\section{Related Work}
\label{sec:related work}

Developing embodied agents that can act and react with life-like motions is essential for creating immersive experiences in games and virtual reality applications. This capability is also of vital importance in robotics, where naturalistic behaviors can improve safety, energy efficiency~\citep{escontrela2022adversarial}, and the learning of general-purposes skills from human data~\citep{Grauman_2022_CVPR}. 
Given the expressive and nuanced nature of human motion, data-driven approaches have emerged as an effective paradigm for producing realistic, life-like behaviors by learning from reference motion data. \emph{Kinematics-based methods} typically train models, such as autoregressive models~\citep{holden2016deep, zhang2018mode, holden2017phase}, variational autoencoders (VAEs)~\citep{ling2020character,rempe2021humor, starke2024categorical}, or diffusion models~\citep{zhang2022motiondiffuse, tevet2023human, shi2024interactive}, on large motion datasets to synthesize plausible character animations. However, most of these methods do not explicitly enforce physical constraints, often leading to physically implausible behaviors, especially for new scenarios and tasks.

\paragraph{Physics-Based Methods}
In contrast, \emph{physics-based methods} aim to create control policies that generate motion within simulated environments, governed by dynamic equations and realistic physical laws~\citep{wampler2014generalizing, raibert1991animation}. 
These controllers are often constructed via trajectory optimization techniques or reinforcement learning (RL)~\citep{peng2017deeploco, wang2009optimizing}. Previous data-free approaches based on heuristics, such as SIMBICON~\citep{yin2007simbicon},
can achieve mechanically functional behaviors, but often produce unnatural motions. Manually designing heuristics that capture the expressiveness of real human movement remains a significant challenge~\citep{faloutsos2001composable, hodgins1995animating, witkin1988spacetime, 2010-TOG-gbwc}. 
Instead of relying on hand-crafted heuristics, data-driven approaches based on model predictive control (MPC) or trajectory optimization have shown promise in emulating naturalistic human motions~\citep{sharon2005synthesis, muico2009contact, lee2010data, liu2010sampling}.
More recently, reinforcement learning-based motion tracking methods have enabled controllers to effectively imitate a wide range of reference motion clips, resulting in more life-like and agile behaviors~\citep{peng2018deepmimic, liu2017learning, won2017train}. However, tracking-based methods limit a controller to closely following a given motion clip, which can impede the controller's ability to generalize to new tasks, particularly when task objectives are not closely aligned with the reference motions. 
To address this limitation, many systems incorporate task-specific motion planners~\citep{bergamin2019drecon, liu2012terrain, park2019learning}, or high-level policies~\citep{yao2022controlvae, yao2024moconvq, zhu2023neural}, which select appropriate reference motions or latent-space skills for the controller to perform in order to complete a given task.

Distribution matching offers an alternative to motion tracking by training controllers to imitate the broader behavioral distribution of a motion dataset rather than tracking reference motions frame-by-frame. Generative Adversarial Imitation Learning (GAIL) approximates this objective using a learned discriminator to distinguish agent's motions from dataset motions~\citep{ho2016generative}. Adversarial Motion Priors (AMP) extend this idea to model flexible style objectives, enabling controllers to produce life-like behaviors for novel tasks that are not observed in the original dataset~\citep{peng2021amp}. However, adversarial priors must be trained jointly with a specific policy for each new task, often requiring retraining and persistent access to the dataset. In contrast, our method achieves comparable performance to state-of-the-art adversarial imitation learning approaches by introducing a reusable and modular score-matching motion prior, which can even be shaped into stylistic priors beyond those contained in the original dataset.

\paragraph{Diffusion Models for Control}

Diffusion models' state-of-the-art performance across a wide range of domain has spurred growing interest in leveraging diffusion models for control. Given their ability to generate high-fidelity motion, a straightforward application is to use task-oriented diffusion models as motion planners~\citep{xu2025parc, tevet2024closd, serifi2024robot, ren2023insactor}. These methods leverage auto-regressive motion diffusion models to predict target future trajectories, which are then executed by a low-level tracking controller. In addition to their use as motion planners, diffuse models have also been used to directly model controllers, enabling controllers to produce flexible multi-modal behaviors for tasks such as manipulation~\citep{chi2023diffusionpolicy, janner2022planning}, and character control~\citep{huang2025diffuse, truong2024pdp, wu2025uniphys, Huang2024DiffuseLocoRL}. 
Unlike these prior methods that focus on incorporating diffusion models as components of a controller, our work aims to repurpose \emph{pretrained} diffusion models as task-agnostic behavioral priors within an optimization objective. Our method utilizes diffusion models as a reward function that can be integrated into general reinforcement learning frameworks to guide policy learning via score distillation sampling.

\paragraph{Score Distillation Sampling}
Pretrained diffusion models are not only capable of generating samples via a denoising process, but can also serve as optimization objectives through score distillation sampling (SDS)~\citep{poole2022dreamfusion}. SDS has enabled pretrained image and video diffusion models to be adapted for a wide range of downstream generative modeling tasks~\citep{wang2023prolificdreamer,Jiang2024Animate3D,Yin_2024_CVPR}.
Inspired by the success of SDS in the vision domain, recent works have also explored incorporating diffusion-based priors into reinforcement learning frameworks for control. A series of methods replace the discriminator in a GAIL-style setup with a diffusion model~\citep{Wang2024DiffAIL, NEURIPS2024DiffRewardAIL, Pang2025DPR, NEURIPS2024DIO}, leveraging its expressive capabilities to differentiate between real and fake trajectories. However, these methods still rely on adversarial training and typically require continual updates to the diffusion model during policy optimization.
Beyond adversarial frameworks, other systems have attempted to apply SDS-like techniques directly to policy learning. For example, \citet{luo2024textaware} guide policy training using pretrained image or video diffusion models conditioned on text prompts. While these methods can produce behaviors that  align with textual instructions, the resulting motions often appear unnatural.
The work that is most reminiscent to ours is SMILING~\citep{wu2025diffusing}, which train controllers using a score-matching objective similar to variational score distillation (VSD) ~\citep{wang2023prolificdreamer}. However, unlike SMILING, which requires training task-specific diffusion models, our method trains a task-agnostic motion prior from unstructured motion data. This prior can then be reused to train diverse control policies by combining it with separate task objectives in a goal-conditioned reinforcement learning framework. We further introduce several key design decisions that enable high-quality naturalistic motions across a diverse repertoire of tasks.

\section{Background}
Our system leverages reinforcement learning to train physics-based control policies and diffusion models to construct score-matching motion priors used as training objectives. In this section, we review the core concepts of reinforcement learning and diffusion models that underpin our method.

\subsection{Reinforcement Learning}
\label{sec:rl}
Our physics-based controllers are trained through a reinforcement learning (RL) framework, where an agent interacts with an environment according to a policy $\pi$ in order to optimize a given objective $J(\pi)$~\citep{sutton1998reinforcement}. At each time step $t$, the agent observes the current state $\rvs_t$, then samples and executes an action according to the policy $\rva_t \sim \pi(\rva_t | \rvs_t)$. The environment then transitions to a new state according to the dynamics $\rvs_{t+1} \sim p(\rvs_{t+1} | \rvs_t, \rva_t)$, and the agent receives a scalar reward $r_t = r(\rvs_t, \rva_t, \rvs_{t+1})$. The objective is to learn a policy that maximizes the expected discounted return:
\begin{equation}
J(\pi) = \mathbb{E}_{\tau \sim p(\tau | \pi)} \left[ \sum_{t=0}^{T-1} \gamma^t r_t \right],
\end{equation}
where $\tau = \{\rvs_0, \rva_0, r_0, \rvs_1, \dots, \rvs_{T-1}, \rva_{T-1}, r_{T-1}, \rvs_T\}$ is a trajectory produced by the policy $\pi$, and $\gamma \in [0,1]$ is a discount factor. The reward function serves as a flexible interface for specifying task objectives. However, crafting effective rewards that consistently induce naturalistic behaviors can be challenging and often requires extensive manual tuning.

\subsection{Diffusion Models and Score Distillation Sampling}
\label{sec:diffusion_models}

In score-based generative modeling methods~\citep{song2019generative, song2021scorebased}, a neural network $f: \sR^D \rightarrow \sR^D$ is trained to estimate the \emph{score} of a point $\rvx$ under a given distribution $p(\rvx)$, which is defined as the gradient of the log-density $\nabla_{\rvx} \log p(\rvx)$. However, the predicted scores may be unreliable, as the available training data typically do not cover the entire space $\sR^D$. In regions with low data density, score estimates tend to be inaccurate. To address this issue, \citet{vincent2011connection} and \citet{song2021scorebased} propose adding noise to the data. When the noise level is sufficiently large, the perturbed data distribution covers the entire space, enabling the training of more robust and scalable score estimators. This principle forms the foundation of our approach. Score-based generative modeling with multi-level noise perturbation allows the construction of reusable motion priors from relatively limited data that lies in a low-dimensional manifold.

\paragraph{Diffusion Models}
Diffusion models approximate a data distribution by progressively corrupting and subsequently denoising samples through a sequence of transformations~\citep{ho2020denoising}. Given a clean sample from the data distribution, the forward process iteratively adds Gaussian noise over $N$ steps to form a Markov chain $\{ \rvx^i \}_{i=0}^{N}$, defined as:
\begin{equation}
    q\left(\rvx^i \mid \rvx^{i-1}\right) = \mathcal{N}\left(\rvx^i; \sqrt{1 - \beta_i} \rvx^{i-1}, \beta_i \mathbf{I}\right),
\end{equation}
where $i$ indicates the noise level, $\{\beta_i\}_{i=1}^N$ is a predefined noise schedule. Since additive \emph{i.i.d.} Gaussian noise is used in the diffusion process, $\rvx^i$ can be conveniently sampled from $\rvx^0$ directly via:
\begin{equation}
\label{eq:ddpmforward}
    \rvx^i =  \sqrt{\bar{\alpha}_i} \rvx^0 + \sqrt{1 - \bar{\alpha}_i} \boldsymbol{\epsilon} ,
\end{equation}
without performing the full iterative diffusion process. Here, $\bar{\alpha}_i = \prod_{j=1}^i (1 - \beta_j)$ and $\boldsymbol{\epsilon} \sim \mathcal{N}(0, \mathbf{I})$. This forward diffusion process resembles the multi-level noise perturbation used in score-based generative models~\citep{song2021scorebased}.
To sample from the learned distribution $p(\rvx)$, a denoising network $f$ is typically trained using the \emph{simple} DDPM objective~\citep{ho2020denoising}:
\begin{equation}
\label{eq:ddpmloss}
    \mathcal{L}_{\text{simple}} = \mathbb{E}_{i, \rvx^0, \boldsymbol{\epsilon}} \left[ \left\| \boldsymbol{\epsilon} - f\left(\rvx^i \right) \right\|^2_2 \right],
\end{equation}
where $\rvx^i$ is generated through the forward diffusion process detailed in \Cref{eq:ddpmforward}. 

\begin{figure}[t]
    \centering
    \includegraphics[width=\linewidth]{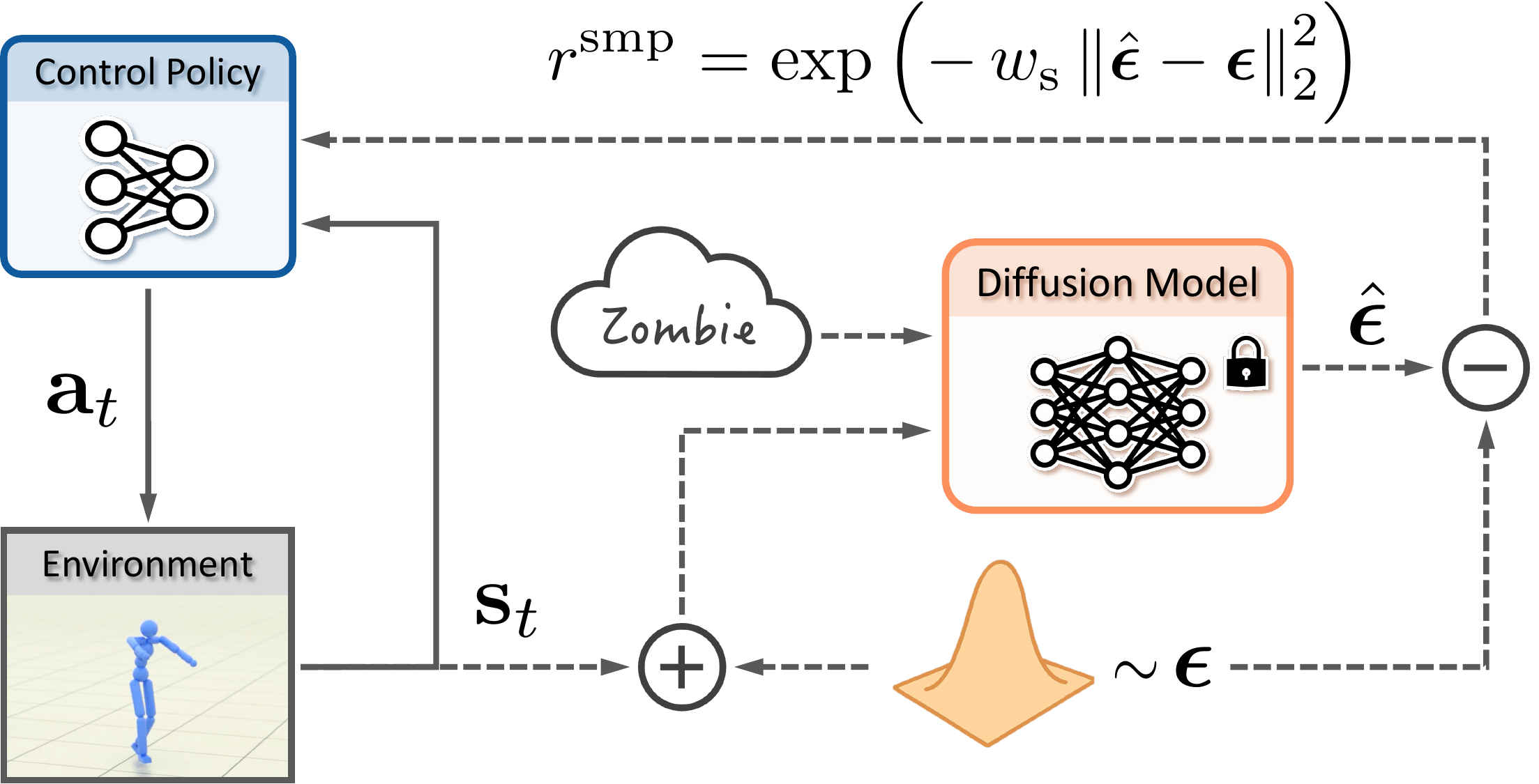}
    \caption{
    Schematic overview of the system. The dashed arrows indicate components used only during policy training. A pretrained motion diffusion model serves as a reusable reward model for motion naturalness via score distillation sampling. The model can be style-conditioned, enabling the policy to learn specific skills or styles without retraining or continuous access to the original motion data. 
    }
    \label{fig:pipeline}
\end{figure}

\paragraph{Score Distillation Sampling (SDS)}
Once a diffusion model $f$ is trained, samples can be generated by applying the reverse diffusion process or through score distillation sampling (SDS)~\citep{poole2022dreamfusion}. SDS minimizes the KL divergence between the distribution of diffused samples derived from the forward diffusion process $q(\rvx^i \mid \rvx^{0})$ and the reference data distribution learned by the pretrained diffusion model $f(\rvx^i)$. 
The gradient of the KL divergence can be estimated using the difference between the score from the forward diffusion process and the prediction from the diffusion model:
\begin{equation}
    \nabla \mathcal{L}_{\mathrm{SDS}}
    = \mathbb{E}_{i, \boldsymbol{\epsilon}} \left[ 
        w(i) \left( f\left(\rvx^i\right) - \boldsymbol{\epsilon} \right) 
        \nabla \rvx
    \right],
\end{equation}
where $\rvx$ is the sample being optimized, and $w(i)$ are coefficients determined by the diffusion noise schedule. 
Previous work has shown that the weighting function $w(i)$ has limited impact and can be substituted with uniform weighting without negatively affecting performance~\cite{threestudio2023}. The SDS gradient can also be derived from the diffusion loss in~\Cref{eq:ddpmloss}, omitting the Jacobian with respect to the score estimator $f$. The SDS loss can be simplified as 
\begin{equation}
 \mathcal{L}_{\mathrm{SDS}}
    =  \left\| \hat{ \boldsymbol{\epsilon}} - \boldsymbol{\epsilon} \right\|_2^2 ,
\end{equation}
where  $\hat{ \boldsymbol{\epsilon}}$  denotes the noise predicted by the denoising network $f$. The optimum of the SDS objective corresponds to samples that minimize the diffusion loss, thereby resembling the characteristics of the dataset on which the diffusion model was originally trained.

\begin{figure}[t]
\centering
\includegraphics[width=\linewidth]{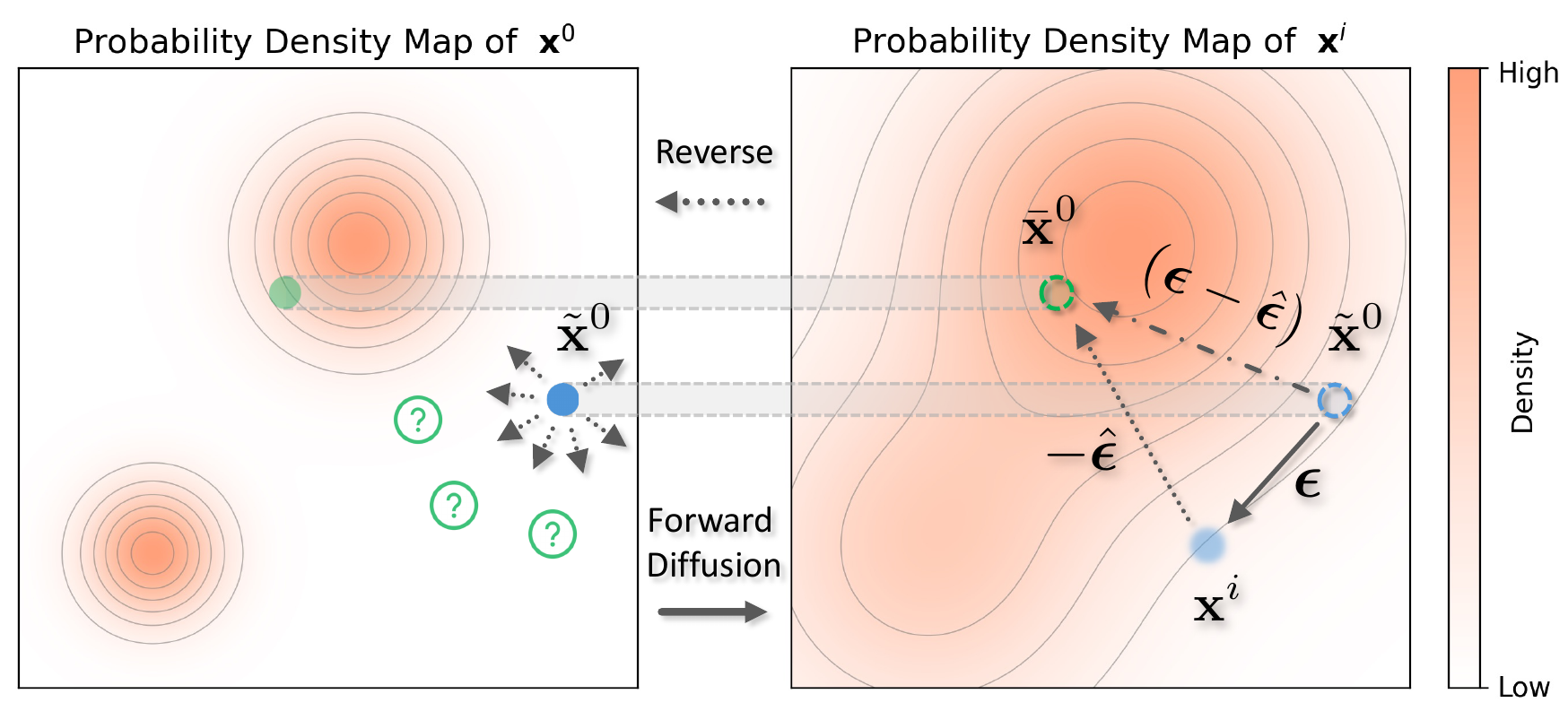}
\caption{
A qualitative illustration of the probability density maps of $\rvx^0$ and $\rvx^i$. When the agent’s motion $\tilde{\rvx}$ deviates significantly from the reference distribution, the gradient of $p(\rvx^0)$ cannot be reliably approximated in low-density regions. The forward diffusion process in \Cref{eq:ddpmforward} maps $\tilde{\rvx}$ to a diffused sample $\rvx^i$, where the density $p(\rvx^i)$ is higher and the score estimate is more reliable. This estimated score can then be used to obtain a pseudo target $\bar{\rvx}^0$ from the reference distribution via a reverse process. The residual between the predicted noise $\hat{\boldsymbol{\epsilon}}$ and the added noise $\boldsymbol{\epsilon}$ provides the correction that aligns the agent’s motion $\tilde{\rvx}$ with the reference distribution.
}
\label{fig:ssmp}
\end{figure}

\section{Overview}
In this work, we introduce the \emph{Score-Matching Motion Prior (SMP)}, a \emph{reusable}, \emph{modular} imitation objective derived from a pretrained diffusion model via score distillation sampling (SDS). The pretrained diffusion model is used to estimate the gradient of the log-likelihood (\emph{i.e.}, the \emph{score}) of the reference distribution, which evaluates the similarity between an agent’s motions and motions in a reference dataset. The robustness of score-based generative modeling enables a pre-trained diffusion model to provide reliable guidance even for motions that are different from those in the original motion dataset. The general style imitation objective provided by SMP can be combined with task-specific objectives to train control policies that can accomplish diverse tasks using natural life-like behaviors.

\Cref{fig:pipeline} illustrates an overview of our framework. 
Unlike prior approaches that use diffusion models as \emph{planners}~\citep{serifi2024robot, tevet2024closd, ren2023insactor}, 
SMP repurposes a pretrained diffusion model as a general \emph{reward model} for evaluating motion naturalness to guide training of a control policy.
The task-agnostic diffusion model $f$ is trained solely on reference motion data, independently of any control policy or task. Once trained, it is frozen and utilized as a reward function via score distillation sampling (SDS). When training a policy, the SDS objective encourages the policy to minimize the discrepancy between the noise $\boldsymbol{\epsilon}$ added to the simulated motion, and the noise $\boldsymbol{\hat{\epsilon}}$ estimated by the pretrained diffusion model. 
The SDS error is minimized when the agent's motions closely aligns with the reference distribution. Similar to other distribution-matching objectives, SMP does not require the simulated character to exactly replicate specific reference motions. Instead, it encourages behaviors that capture the general characteristics of the reference data, enabling smooth transitions and adaptation to tasks that may require skills not explicitly present in the dataset. Furthermore, SMP can be trained on large and diverse datasets by employing a conditional diffusion model. This enables policies to acquire different stylistic behaviors by conditioning the pretrained diffusion model on style labels, without the need to retain the original motion dataset.

\section{Score-Matching Motion Priors}
\label{sec:method}

A score-matching motion prior is modeled as a diffusion model, which is trained to predict the \emph{score} of noisy input motions. The predicted score can be interpreted as a correction applied to a noisy sample to denoise back to a clean sample from the training data distribution. This enables SMP to construct motion imitation objectives directly from the pretrained motion diffusion model.

Given a motion dataset, we first train a motion diffusion model to generate motion clips consisting of $H$ consecutive frames $\rvx := (\rvs_{t-H+2}, \dots, \rvs_{t+1})$. During policy training, each simulated motion clip produced by the agent is diffused with Gaussian noise $\boldsymbol{\epsilon} \sim \mathcal{N}(0, \mathbf{I})$ to a diffusion timestep $i$ through the forward diffusion process in \Cref{eq:ddpmforward}, resulting in a noisy sample $\rvx^i$. The pretrained diffusion model $\hat{\boldsymbol{\epsilon}} = f(\rvx^i)$ then predicts the score $\hat{\boldsymbol{\epsilon}}$ at $\rvx^i$, which provides a denoising direction back toward the reference motion distribution. The correction that should be applied to align the agent’s motion with the reference distribution is therefore given by the noise residual $(\boldsymbol{\epsilon} - \hat{\boldsymbol{\epsilon}})$, as illustrated in \Cref{fig:ssmp}. The SMP reward used for motion imitation is then defined as
\begin{equation}
    r^{\mathrm{smp}} = \exp\left(-\, w_\mathrm{s}\left\| \hat{\boldsymbol{\epsilon}} - \boldsymbol{\epsilon} \right\|_2^2\right),
\end{equation}
which is maximized when the simulated motion aligns with the reference distribution. Following standard RL reward design~\cite{peng2018deepmimic}, we apply an exponential transformation to the SDS loss to normalize the reward between $[0,1]$. While this deviates from the original SDS formulation, we find that this normalization produces empirical improvements when training RL policies.

Previous SDS-based methods have primarily shown promise in the 3D generation, but often produce low-quality, ``blurry'' results \citep{liang2024luciddreamer}. 
Earlier efforts to use SDS as a reinforcement learning objective have also struggled to match the performance of adversarial methods for training control policies~\citep{luo2024textaware, wu2025diffusing}. 
In the following sections, we introduce several key design decisions that improve stability for policy training using SMP, enabling policies to learn high-quality naturalistic behaviors.

\begin{figure}[t]
\centering
\includegraphics[width=0.9\linewidth]{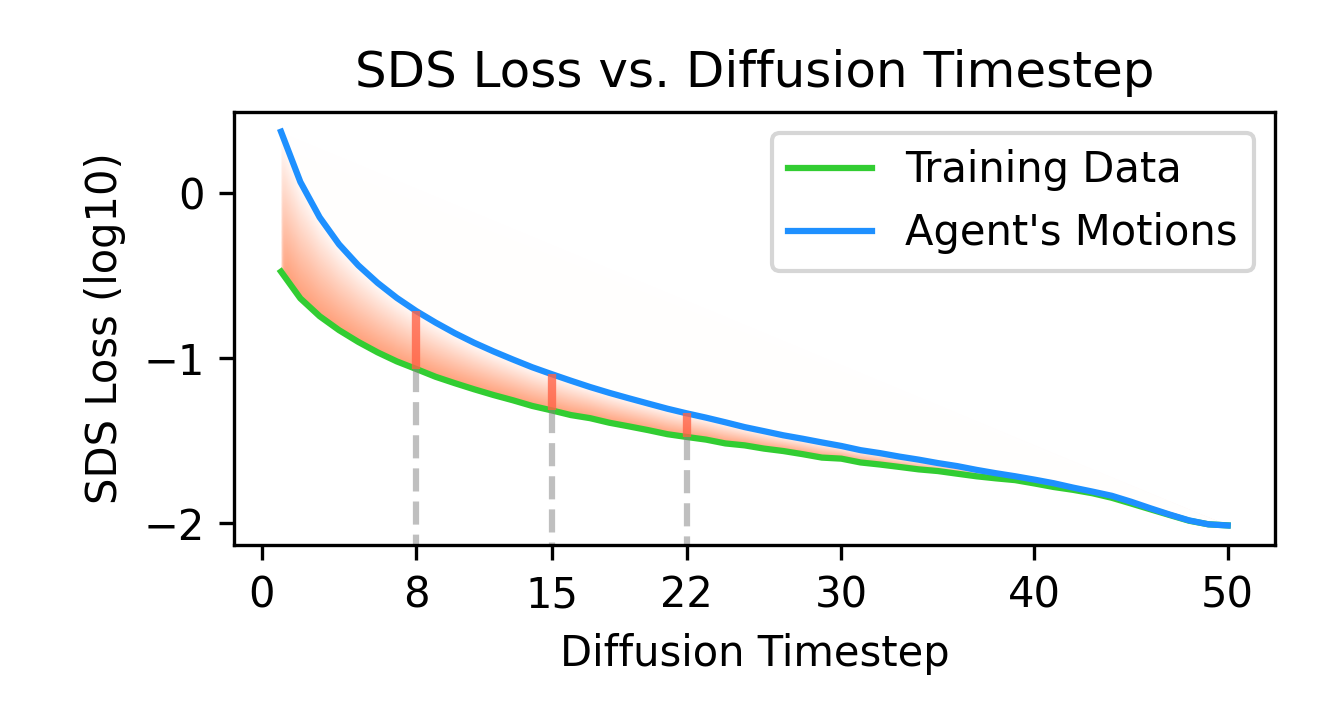}
\caption{
    An example of the SDS loss across diffusion noise levels. The $y$-axis shows $\log_{10}$ values for visualization. Larger diffusion timesteps correspond to higher noise levels applied to the sample. Statistics are averaged across 1024 samples for each diffusion timestep. The orange region highlights the difference between the SDS scores from the agent's motions and the motion data. The scale of the SDS loss can vary significantly across different diffusion noise levels.
    }
\label{fig:sdsvstimestep}
\end{figure}

\subsection{Ensemble Score-Matching}

The SDS objective can be highly sensitive to the choice of diffusion timestep $i$, as different noise scales provide disparate forms of guidance~\citep{lin2023magic3d}.  
At higher noise levels, diffused samples are heavily corrupted and dominated by Gaussian noise, which matches the training settings of the diffusion model and leads to more reliable score predictions due to less susceptibility to out-of-distribution samples. Therefore, evaluating the objective at higher diffusion timesteps $i$ can be more instructive when the agent's motions deviate substantially from the reference data distribution. 
However, excessive noise also removes much of the information present in the agent behaviors, which may prevent the objective from correcting more minute errors in the agent's motions. In such cases, the guidance signal may encourage the policy to produce generic or averaged behaviors, as the model no longer retains sufficient information to steer toward more realistic and detailed motions.
Therefore, when the difference between the agent's motions and motions in the dataset is relatively small, diffusing the sample to lower noise levels can better preserve information of the agent's motions and apply finer-grained corrections.

In prior SDS-based methods in vision domains~\citep{poole2022dreamfusion, threestudio2023}, it is common to sample the diffusion noise level uniformly $i \sim \mathcal{U}(1, N)$. However, in reinforcement learning, this stochasticity can introduce undesirable variance into the reward signal, leading to less reliable value and advantage estimation. For example, samples diffused with high noise levels are nearly pure Gaussian noise, allowing the diffusion model to easily predict the applied noise, leading to small SDS errors and limited information about differences between the agent’s motions and the reference distribution, as illustrated in \Cref{fig:sdsvstimestep}. Whereas samples with lower noise levels generally produce larger SDS errors. Due to this noise-dependent variation in the SDS loss, computing the SMP reward with randomly sampled timesteps can result in a highly stochastic and unreliable measure of the similarity between the agent’s motions and the reference data distribution.

To reduce the variance of the SMP reward function, instead of randomly sampling a single timestep per reward calculation, we propose \emph{ensemble score-matching} (ESM), which constructs a lower variance reward function by aggregating multiple SDS evaluations over a fixed set of diffusion timesteps $i \in \sK$. The resulting SMP reward is given by:
\begin{equation}
\label{eq:averageprior}
r^{\mathrm{smp}} = \exp\left(- \frac{w_s}{|\sK|} \sum_{i \in \sK}\left\| \hat{\boldsymbol{\epsilon}}_i - \boldsymbol{\epsilon}_i \right\|_2^2 \right),
\end{equation}
where $\hat{\boldsymbol{\epsilon}}_i = f\left(\sqrt{\bar{\alpha}_i}\,\tilde{\rvx}^0 + \sqrt{1 - \bar{\alpha}_i}\,\boldsymbol{\epsilon}_i\right)$, $\tilde{\rvx}$ denotes the simulated character's motion, and  $\sK$ is a predefined set of diffusion timesteps.
When the diffusion model’s predictions are sufficiently reliable, \emph{i.e.}, for samples that are less out-of-distribution (OOD), the SDS loss computed at lower noise levels responds more sensitively to discrepancies between the agent’s motion and the reference training distribution. 
In our experiments, we found that SDS error computed at low diffusion timesteps is highly sensitive to jittery artifacts, whereas SDS error computed at medium timesteps is more robust and provides better guidance for correcting behavioral artifacts. Moreover, evaluating a large number of diffusion steps incurs significant computational cost, and adjacent timesteps tend to provide redundant guidance at similar granularity levels. We therefore found that evaluating the SMP objective using a small set of representative diffusion timesteps is sufficient for effective performance. In all experiments, we compute the SDS loss at $\sK=\{22,15,8\}$, which provides an effective balance across different noise scales.

In addition, we apply adaptive normalization using the running mean $\mu_i$ of the SDS error at each diffusion timestep $i$ to mitigate the varying loss scales at different noise levels. This adaptive normalization also reduces SMP's sensitivity to variations across different pre-trained diffusion models and behavioral styles, thereby reducing the need for manual parameter tuning for the SMP reward function.

\subsection{Generative State Initialization}

Reference state initialization (RSI), where the simulated character is initialized to states randomly sampled from a reference motion dataset, has been shown to be a vital technique for improving exploration in motion imitation methods~\citep{peng2018deepmimic}. However, RSI requires access to the motion dataset to sample initial states from during policy training. To remove this reliance on the original dataset, SMP can instead leverage the generative prior to \emph{generate} initial states when training the policy. When using SMP to train a policy, initial states are generated by sampling from the diffusion model used as the pretrained motion prior. SMP therefore serves dual purposes in our framework, as both a reward function and an initial state distribution when training new policies. This \emph{generative state initialization} (GSI) method alleviates the need to retain the original motion dataset once the SMP has been trained by leveraging the generative capabilities of diffusion models to produce diverse, high-quality initial states.

\section{Model Representation}

Our framework utilizes a pretrained diffusion model as a reusable, modular motion prior for imitation learning. The task-agnostic motion diffusion model serves as a reward function that evaluates the naturalness of the agent’s motion within a reinforcement learning framework. This allows for the training of control policies that accomplish diverse tasks while exhibiting naturalistic behaviors that resemble those in the original motion dataset. In this section, we detail key design decisions of the learning framework.

\begin{algorithm}[t!]
\caption{Policy Training with SMP}
\label{alg:SMP}
\begin{algorithmic}[1]

\STATE{\textbf{Input (optional):} style label $c$ }

\STATE{$f \gets$ load pretrained diffusion model}
\STATE{$\pi \gets$ initialize policy}
\STATE{$V \gets$ initialize value function}
\STATE{$\mathcal{B} \gets \emptyset$ \ initialize reply buffer}

\item[]
\WHILE{not done}
    \FOR{trajectory $j = 1,...,m$}
    	\STATE{$\tau^j \gets \{(\rvs_t, \tilde{\rvx}_t, \rva_t, r^g_t)_{t=0}^{T-1}, \ \rvs^g_T, \tilde{\rvx}_T\}$ collect trajectory with $\pi$}
        \FOR{$t = 0,...,T-1$}
                \FOR{diffusion timestep $i \in \sK$}
                    \STATE{$\boldsymbol{\epsilon}_i \sim \mathcal{N}(0, I)$}
                    \STATE{$\hat{\boldsymbol{\epsilon}}_i \gets f\left(\sqrt{\bar{\alpha}_i}\,\tilde{\rvx}_{t+1} + \sqrt{1 - \bar{\alpha}_i}\,\boldsymbol{\epsilon}_i, \, c\right)$}
                \ENDFOR
                \STATE{$r^\mathrm{smp}_t \gets$ compute prior reward via \cref{eq:averageprior} using $\{\boldsymbol{\epsilon}, \hat{\boldsymbol{\epsilon}}\}^\sK_i$}
            \STATE{$r_t \gets w^\mathrm{prior} r^\mathrm{smp}_t + w^g r^g_t$}
            \STATE{record $r_t$ in $\tau^j$}
        \ENDFOR
        \STATE{store $\tau^j$ in $\mathcal{B}$}
    \ENDFOR
    
    \item[]
    \STATE{update $V$ and $\pi$ using data from trajectories $\{\tau^j\}_{j=1}^m$}
        
\ENDWHILE
\end{algorithmic}
\end{algorithm}

\subsection{Motion Representation}
\label{subsec:motion_rep}
Designing an appropriate motion representation is critical both for training the diffusion model to capture the dataset distribution, and for repurposing it as a motion prior for policy training. The representation should contain sufficient information to reconstruct motion while remaining easy to extract from both kinematic motion clips and simulator state observations. Following the design of prior works in motion diffusion and AMP~\citep{tevet2023human, zhang2022motiondiffuse,peng2021amp}, our motion features include:
\begin{itemize}
    \item Root linear and angular velocities, represented in the character’s local coordinate frame at the last timestep in a motion segment.
    \item Local joint rotations.
    \item 3D positions of end-effectors (\emph{e.g.}, hands and feet), represented in the character’s local coordinate frame.
\end{itemize}
The character's local coordinate frame is defined with the origin at the root (\emph{i.e.}, pelvis), the $x$-axis aligned with the root link’s facing direction, and the $y$-axis aligned with the global up vector. Joint rotations are represented using a 6D representation for spherical joints~\citep{zhou2019continuity}.

\subsection{Diffusion Model Implementation}

The motion diffusion model is implemented as a transformer encoder, using adaptive normalization to inject noise-level conditions (and style conditions, when available). Unless otherwise specified, we use a window of $H=10$ frames. We find that a simple two-layer transformer encoder with only 3M parameters is sufficient to capture the distribution of large-scale motion datasets, such as 100STYLE~\citep{mason2022real} with over 20 hours of stylized motions. The number of diffusion timesteps is set to $N=50$, and the model is trained to predict $\boldsymbol{\epsilon}$. Following standard practice for training diffusion models, we apply exponential moving average (EMA) on the model parameters during training~\citep{karras2024analyzing, dhariwal2021diffusion}.
The model is trained for 400k–800k iterations, depending on the dataset, with convergence typically achieved within five hours on a single RTX 4090 GPU.

\subsection{Model Implementation}
Following prior work, the control policy $\pi$ is modeled using a multi-layer perceptron (MLP) that maps an input state $\rvs_t$ and goal $\rvg$ to a Gaussian distribution over the action space $\rva_t \in \mathcal{A}$~\citep{peng2018deepmimic}. The value function $V(\rvs_t, \rvg)$ is modeled by a separate MLP network with a similar architecture to the policy.

\paragraph{States and Actions}
The state $\rvs_t$ is represented using features similar to those in \citet{peng2021amp}, such as body link positions, link rotations encoded in a 6D representation, and linear and angular velocities of each link. All features are calculated in the character’s local coordinate frame. Since our policies are not trained to track specific reference motions, no phase variable or target reference state information is provided in the state. The action $\rva_t$ specifies target joint positions for PD controllers at each joint. For spherical joints, each target rotation is parameterized using a 3D exponential map $q \in \mathbb{R}^3$~\citep{grassia1998practical}.

\paragraph{Training}
The policy is trained using proximal policy optimization (PPO)~\citep{schulman2017proximal}. At each timestep $t$, the agent queries the motion prior (see \Cref{alg:SMP}) to obtain a prior reward $r_t^\mathrm{smp}$, and may also receive a task reward $r_t^g$ from the environment. These are combined linearly to form the composite reward at that timestep \begin{equation}
    r_t = w^{\mathrm{prior}} r_t^\mathrm{smp} + w^g r_t^g, 
\end{equation}
following \citet{peng2021amp}.  After collecting a batch of trajectories, mini-batches are sampled from the buffer to update both the policy and the value function. The policy is updated with PPO using advantages estimated via $\mathrm{GAE}(\lambda)$~\citep{schulman2015high}, while the value function is trained with targets computed via $\mathrm{TD}(\lambda)$~\citep{sutton1998reinforcement}. The pretrained diffusion model is kept fixed during the policy training process, and does not need to be updated. 

Training a policy with SMP on a single motion clip (e.g., \emph{spinkick}) takes approximately 30 minutes on a single RTX~4090 GPU. Training a HighKnees policy for the target location task takes 11.5 hours for 600M samples, compared to 6.2 hours for AMP. SMP uses a larger prior model that captures 100 styles, whereas AMP uses a smaller style-specific discriminator. Runtime performance is identical to AMP, since only the policy is executed.

\begin{figure*}[t]
	\centering
    \subfigure[Dodgeball \label{fig:dodge}]{\includegraphics[width=2.08\columnwidth]{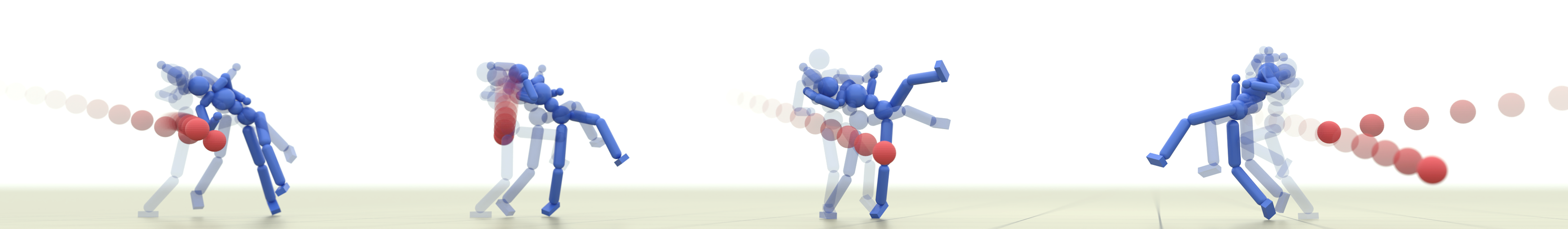}}\\
    \subfigure[Target Location \label{fig:lafan_loc}] {\includegraphics[width=1.04\columnwidth]{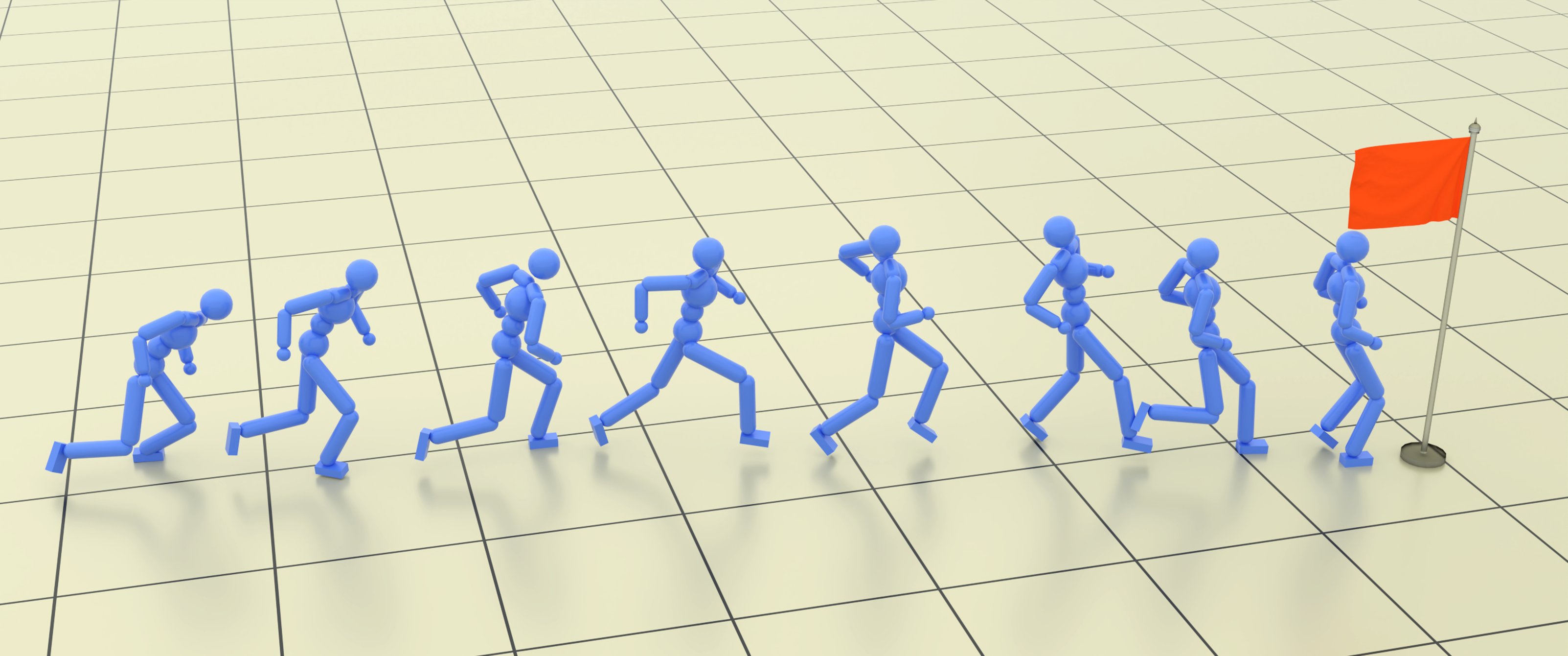}}
    \subfigure[Steering \label{fig:heading}]{\includegraphics[width=1.04\columnwidth]{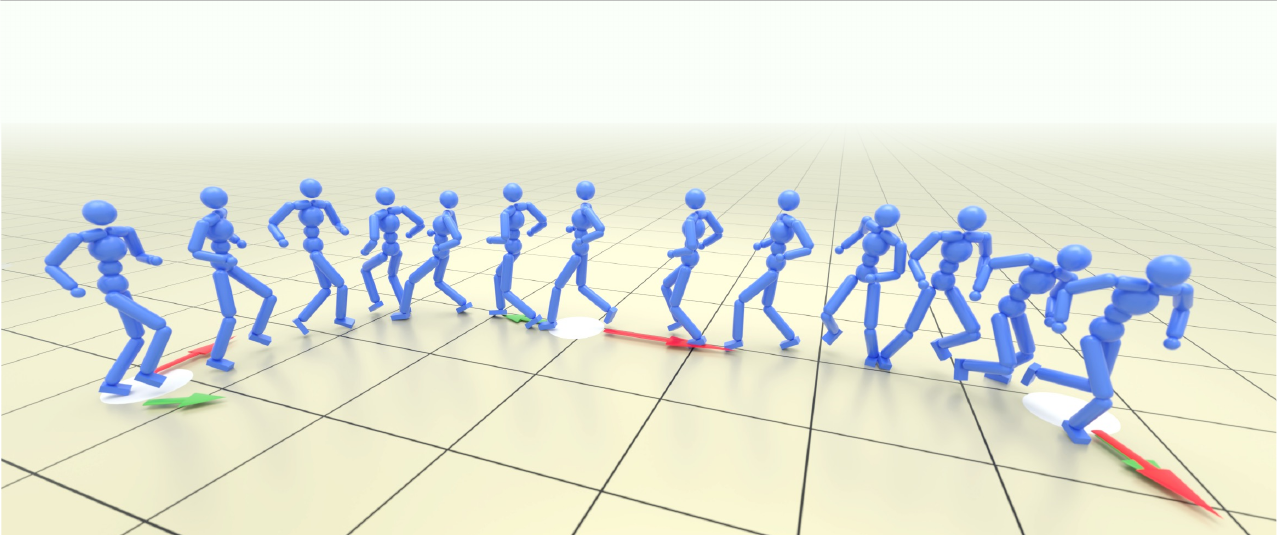}}\\
    \subfigure[Target Speed \label{fig:speed}]{\includegraphics[width=1.04\columnwidth]{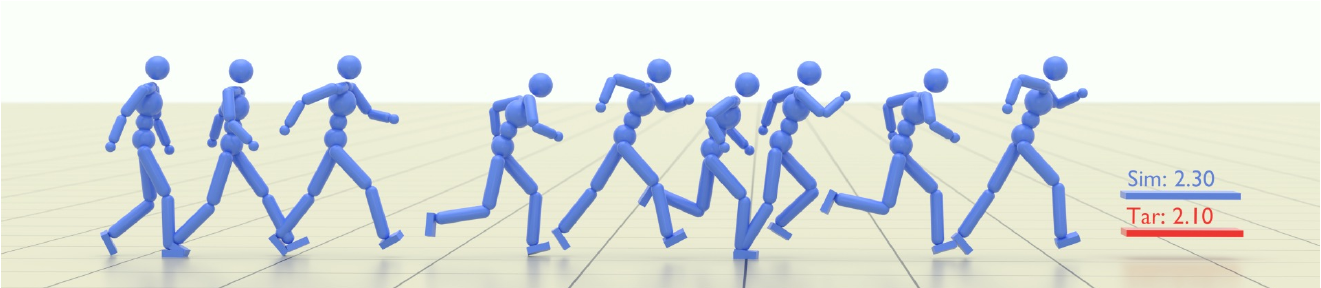}}
    \subfigure[Getup \label{fig:getup1}]{\includegraphics[width=1.04\columnwidth]{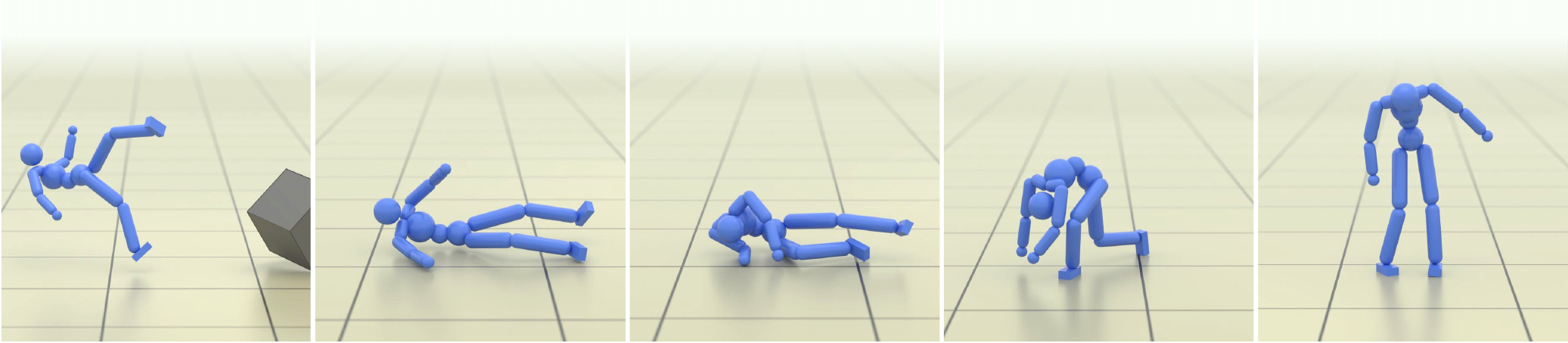}}\\
    \subfigure[Object Carry \label{fig:hoi}]{\includegraphics[width=2.08\columnwidth]{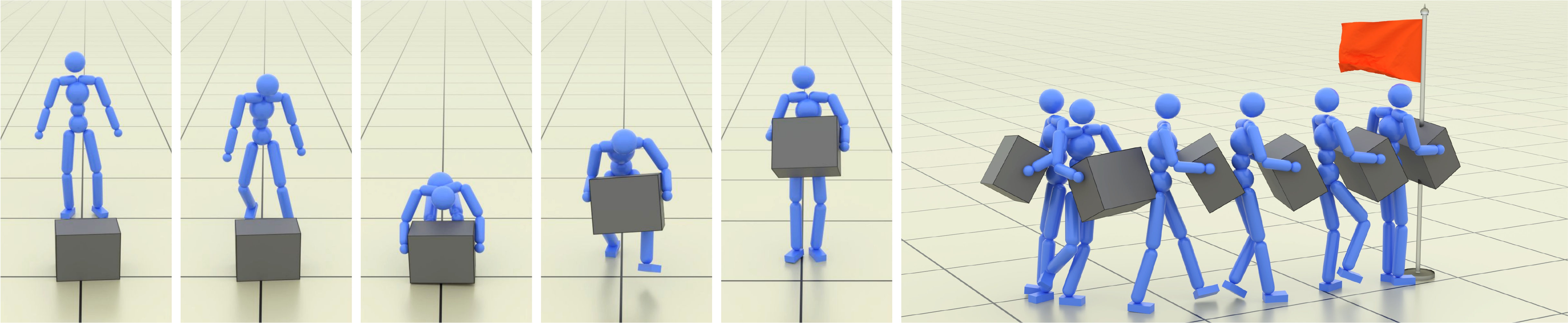}}\\
    \caption{
    Score-matching motion priors can be trained on a wide range of dataset sizes, independently of any task or control policy. Once trained, an SMP provides a motion imitation objective $r^{\mathrm{smp}}$ that can be composed with task rewards $r^{\mathrm{g}}$ to train policies that perform a diverse array of tasks, all the while exhibiting natural, life-like behaviors.
    }
\end{figure*}

\section{Tasks}
\label{sec:task_descrip}
We evaluate the effectiveness of SMP across seven motion control tasks, showcasing its ability to train policies that can perform various control tasks using diverse motion styles. Below, we summarize each task and the corresponding goal observation $\rvg$, which provides the agent with task-relevant information from the environment. More comprehensive details, including task reward functions and specific hyperparameter settings are provided in \Cref{app:task_implementation}.

\paragraph{Target Speed} 
This task requires the character to move at a target speed. The goal for the policy is specified as  $\rvg_t = v_t^*$. The target speed $v_t^*$ is randomly sampled from $[1.2, 6.8]\mathrm{m/s}$. To focus on speed tracking and gait transitions induced by different speeds, the target movement direction is kept fixed.

\paragraph{Steering} The task requires the character to face a specified 2D heading direction $\rvh_t^*$ while simultaneously traveling at a target speed $v_t^*$ along a target horizontal direction $\rvd_t^*$. The steering policy receives the goal information as $\rvg_t = (\rvd^{*}_t, v_t^*, \rvh_t^*)$.

\paragraph{Target Location} In this task, the character is instructed to reach a 2D target location specified on the floor plane. The agent perceives the goal via $\rvg_t = \rvp_t^*$, where the target location $\rvp_t^*$ is represented in the character’s local coordinate frame.

\paragraph{Dodgeball} 
In this task, a ball is launched toward the character from a random position up to $10\mathrm{m}$ away, with a launch speed sampled from $[20,25]\mathrm{m/s}$. This gives the agent less than $0.5\mathrm{s}$ to react and dodge. If the character is hit, the episode terminates early and the agent receives zero reward for all remaining timesteps as a penalty. The agent observes the ball state through goal observations $\rvg_t = \left(\rvp_t^{\mathrm{ball}}, \dot{\rvp}_t^{\mathrm{ball}}\right)$, where $\rvp_t^{\mathrm{ball}}$ and $\dot{\rvp}_t^{\mathrm{ball}}$ denote the ball’s position and velocity represented in the character’s local coordinate frame.

\paragraph{Object Carry} In addition to training priors for human motion, SMP can also be used to train priors that jointly model the interactions between the character and objects. First, we train an SMP on a dataset of human-object carrying motions. This prior is then used to train policies for an object carrying task, where the character is required to carry a box from the ground to a randomly placed target location. The goal $\rvg_t = {\rvp^{\mathrm{box}}_t}^{*}$ records the target box position. The state $\rvs_t$ is augmented with additional features that describe the state of the box, including the position $\rvp^\mathrm{box}_t$ and the orientation $\rvq^\mathrm{box}_t$. All features are represented in the character’s local coordinate frame.

\paragraph{Stair Traversal} To further evaluate SMP's ability for modeling character-object interactions, we apply SMP to train policies for a stair traversal task. The task requires the character to traverse a short staircase by walking up and then down a series of steps while following a target velocity $\mathbf{v}^{\text{tar}}$. At each timestep, the goal $\mathbf{g}_t=\hat{\mathbf{v}}^{\text{tar}}_t$ records the target velocity expressed in the character's local coordinate frame. The state $\rvs_t$ is augmented with additional features that encode the staircase’s position and orientation.

\paragraph{Getup} In this task, the policy is trained to recover from arbitrary fallen states and maintain a standing pose.

\section{Experiments}
To evaluate the effectiveness of SMP, we apply our framework to train policies for a suite of challenging motion control tasks.
In \Cref{subsec:100style}, we demonstrate the \emph{modularity} of SMP by using a pre-trained prior to train new control policies without requiring access to the original motion dataset.
By training a style-conditioned diffusion model on the 20-hour 100STYLE dataset, we show that a single pre-trained prior can then be used to train policies to perform tasks in a variety of different styles, where each style-specific prior reward is obtained simply by conditioning the pretrained SMP on the desired style label. New styles can also be synthesized by composing the pretrained diffusion model's predictions from different styles.
In \Cref{subsec:lafan}, we show that a single motion prior can be \emph{reused} to train policies across diverse tasks, such as steering, target location, and dodgeball.
In addition to guiding the behavioral style of the character, in \Cref{subsec:hoi} we show that SMP can also model \emph{human-object interactions}. 
By training the diffusion model on human-object interaction data, SMP learns to jointly model both object and character motions, allowing agents to use human-like behavior to perform object interaction tasks. 
Beyond task-directed behaviors, SMP also enables agents to learn robust and natural recovery behaviors, as demonstrated in \Cref{subsec:getup}. 
Similar to prior distribution-matching objectives, the policies trained with SMP can synthesize new skills that are not explicitly present in the reference dataset. 
In \Cref{subsec:wjr}, we show that an SMP trained on a small-scale dataset, containing only 3 seconds of walking, jogging, and running motion clips, can automatically lead to the emergence of different locomotion gaits that enable a character to closely follow a wide range of different target speeds, as well as natural transitions between various gaits.
Finally, to benchmark SMP’s effectiveness for motion imitation, we evaluate SMP on single-clip imitation tasks (\Cref{subsec:singleclip}). SMP is able to closely reproduce a diverse set of dynamic and acrobatic skills, achieving comparable motion quality to state-of-the-art adversarial imitation learning methods.

\begin{figure*}[t]
\centering
\includegraphics[width=\textwidth]{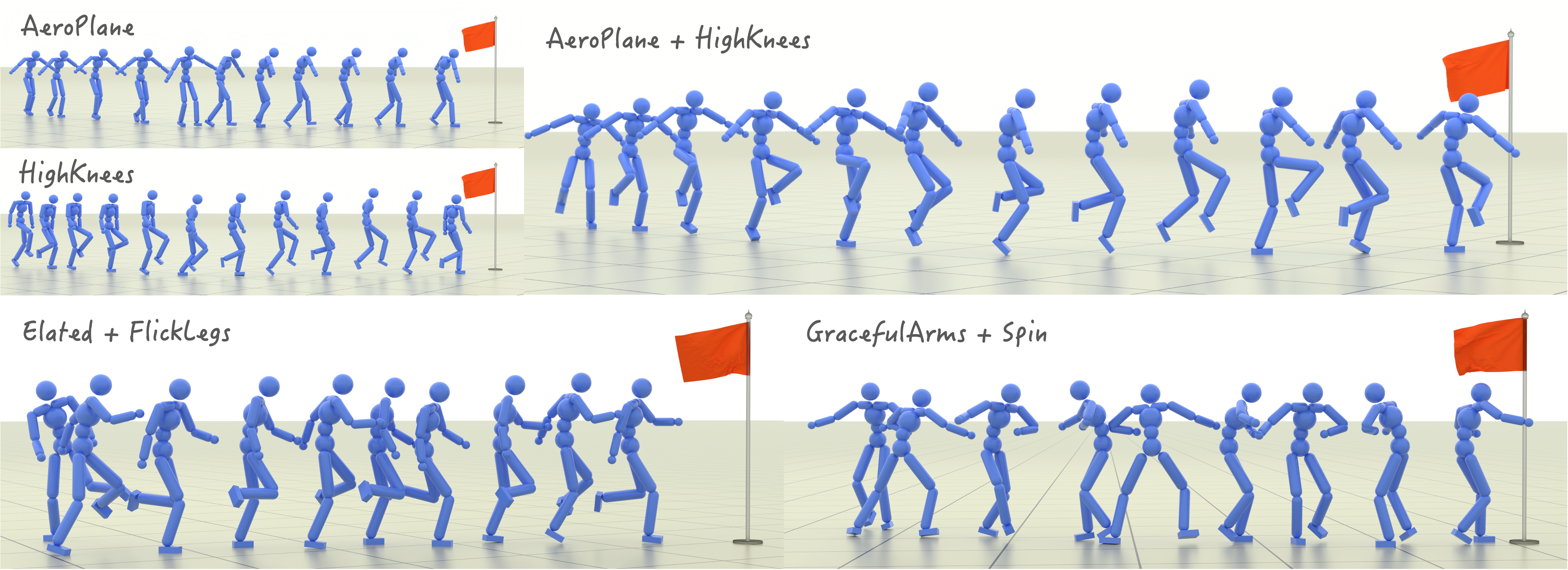}
\caption{
A pretrained 100-style motion prior can be adapted to synthesize motion priors of new styles. For example, we create a novel ``AeroPlane + HighKnees'' prior by blending two existing styles in the $\boldsymbol{\epsilon}$-space.
This crafted prior, which is used for both generative state initialization and the motion prior objective $r^{\mathrm{smp}}$, enables the agent to perform the target location task with a new, agile style, all without requiring the corresponding reference data. 
}
\label{fig:compositionalprior}
\end{figure*}

\paragraph{Baselines:} In the following experiments we compare the performance of SMP with AMP \citep{peng2021amp}, a widely-used adversarial imitation learning method.
Following \citet{peng2021amp}, the reward weights for AMP are set to $w^{\mathrm{prior}} = w^{g} = 0.5$. 
In addition to the standard AMP method, we also compare with a variant of AMP that uses a \textit{frozen discriminator}. This configuration evaluates whether the learned adversarial motion prior can be effectively \emph{reused} without further training. First, a control policy and discriminator are trained jointly using AMP. Then the trained discriminator is reused to train new control policies on the same task without further finetuning using data from the new policy. Furthermore, we include a simple \emph{tabula-rasa} learning baseline (w/o Prior), where the policy is trained solely to maximize the task reward ($w^g = 1$), without any motion priors.

\subsection{One Motion Prior for $100+\mathrm{N}$ Styles}
\label{subsec:100style}
SMP serves as a modular motion-style reward model that guides policy training without requiring access to the original motion dataset. 
This enables the application of various adaptation techniques to the diffusion-based reward model that can further shape the motion prior.
To demonstrate this capability, we train a general 100style-conditioned motion diffusion model $f(\rvx^i, c)$ using the entire 20-hour 100STYLE dataset, where $c$ is a style label. 
Classifier-free guidance (CFG) can then be applied to reshape this general prior into style-specific motion priors~\citep{ho2022classifier}: 
\begin{equation*}
    f_{\mathrm{style}} = f(\rvx^i, \emptyset) + w_\mathrm{cfg} \left(f(\rvx^i, c_\mathrm{style}) - f(\rvx^i, \emptyset)\right),
\end{equation*}
where the style-conditioned prediction $f(\rvx^i, c_{\mathrm{style}})$ is applied to the unconditional prediction $f(\rvx^i, \emptyset)$ according to the guidance weight $w_{\mathrm{cfg}}$.
These adapted priors can be used to train policies for a given task using different behavioral styles. The performance statistics for policies trained with various styles are reported in \Cref{tab:100style}. All policies are trained using the same underlying SMP model conditioned on  different style labels. 
We find that simply setting the CFG scale to 1.0 is generally sufficient to specialize the 100-style prior into distinct style-specific priors, enabling agents to perform tasks with diverse stylistic behaviors, as shown in \Cref{fig:teaser}.
While AMP can produce comparable results, it requires training different style-specific discriminators using curated style-specific datasets.
In contrast, using a fixed pretrained discriminator (AMP-Frozen) is ineffective in producing the desired stylistics behaviors. With AMP-Frozen, we observe that the discriminator’s accuracy drops over the course of policy training, which indicates that the policy is exploiting errors in the discriminator. This exploitation often leads to unnatural behaviors that nonetheless elicit a high score from the discriminator. 
Qualitative comparisons of the various methods are available in the supplementary video.

\begin{table*}[t]
\caption{
Performance of policies trained with different styles on the target location task. Task returns are normalized to $[0,1]$. Style accuracy is evaluated using a style classifier trained on the 100STYLE dataset.
AMP models are trained using style-specific motion datasets, whereas SMP is pretrained once on the full 100STYLE dataset and adapted to each style using CFG during policy training. 
The modularity of SMP enables effective training of style-specific policies without access to the original dataset. AMP with a frozen discriminator is ineffective for producing policies of the desired styles.
}
\centering
\small
\setlength{\tabcolsep}{5pt}
\begin{tabular}{c|c|ccc|ccc}
\toprule
\multirow{3}{*}{\textbf{Dataset}} & \multirow{3}{*}{\textbf{Style}} & \multicolumn{3}{c|}{\textbf{Task Return}} & \multicolumn{3}{c}{\textbf{Style Accuracy}}\\
 & & AMP & \begin{tabular}{c} {AMP} \\[-0.5ex] {Frozen} \end{tabular} & \begin{tabular}{c} {SMP} \\[-0.5ex] {(Ours)} \end{tabular}  & AMP & \begin{tabular}{c} {AMP} \\[-0.5ex] {Frozen} \end{tabular} & \begin{tabular}{c} {SMP} \\[-0.5ex] {(Ours)} \end{tabular} \\
\midrule
\multirow{13}{*}{100STYLE}   & AeroPlane & $0.867^{\pm 0.001}$ & $0.871^{\pm 0.008}$ & $0.882^{\pm 0.010}$ & $0.981^{\pm 0.004}$ & $0.000^{\pm 0.000}$ & $0.995^{\pm 0.003}$\\
                            & Chicken & $0.876^{\pm 0.005}$ & $0.874^{\pm 0.003}$ & $0.877^{\pm 0.008}$ & $0.973^{\pm 0.019}$ & $0.306^{\pm 0.530}$ & $0.989^{\pm 0.019}$ \\
                            & CrossOver & $0.866^{\pm 0.001}$ & $0.470^{\pm 0.363}$ & $0.879^{\pm 0.002}$ & $0.994^{\pm 0.005}$ & $0.320^{\pm 0.554}$ & $0.994^{\pm 0.005}$ \\
                            & Dinosaur & $0.886^{\pm 0.001}$ & $0.856^{\pm 0.020}$ & $0.882^{\pm 0.015}$ & $0.998^{\pm 0.004}$ & $0.004^{\pm 0.006}$ & $0.999^{\pm 0.001}$ \\
                            & FlickLegs & $0.881^{\pm 0.001}$ & $0.580^{\pm 0.269}$ & $0.868^{\pm 0.012}$ & $0.983^{\pm 0.004}$ & $0.009^{\pm 0.011}$ & $0.979^{\pm 0.024}$  \\
                            & HandsBetweenLegs & $0.870^{\pm 0.003}$ & $0.888^{\pm 0.007}$ & $0.850^{\pm 0.004}$ & $0.690^{\pm 0.198}$ & $0.000^{\pm 0.000}$ & $0.978^{\pm 0.007}$ \\
                            & HighKnees & $0.882^{\pm 0.003}$ & $0.872^{\pm 0.014}$ & $0.897^{\pm 0.003}$ & $0.988^{\pm 0.009}$ & $0.012^{\pm 0.017}$ & $0.992^{\pm 0.005}$ \\
                            & Neutral & $0.873^{\pm 0.007}$ & $0.755^{\pm 0.103}$ & $0.891^{\pm 0.016}$ & $0.991^{\pm 0.005}$ & $0.426^{\pm 0.478}$ & $0.999^{\pm 0.001}$ \\
                            & Skip & $0.875^{\pm 0.003}$ & $0.512^{\pm 0.321}$ & $0.891^{\pm 0.008}$ & $0.985^{\pm 0.003}$ & $0.425^{\pm 0.479}$ & $0.646^{\pm 0.350}$ \\
                            & Spin (Clockwise) & $0.871^{\pm 0.001}$ & $0.874^{\pm 0.012}$ & $0.858^{\pm 0.004}$ & $0.990^{\pm 0.003}$ & $0.006^{\pm 0.010}$ & $0.978^{\pm 0.015}$ \\
                            & Superman & $0.872^{\pm 0.004}$ & $0.870^{\pm 0.013}$ & $0.893^{\pm 0.005}$ & $0.998^{\pm 0.004}$ & $0.304^{\pm 0.492}$ & $0.999^{\pm 0.002}$ \\
                            & Zombie & $0.872^{\pm 0.003}$ & $0.823^{\pm 0.039}$ & $0.875^{\pm 0.014}$ & $0.973^{\pm 0.005}$ & $0.649^{\pm 0.563}$ & $0.996^{\pm 0.006}$ \\
\midrule
\multicolumn{2}{c|}{Average} & $0.874$ & $0.771$ & \cellcolor{green!10}$0.879$ & \cellcolor{green!10}$0.962$ & 0.205 & \cellcolor{green!10}$0.962$\\ 
\bottomrule
\end{tabular}
\label{tab:100style}
\end{table*}

\paragraph{Style Composition}
SMP supports crafting novel motion priors by composing the diffusion model’s outputs, enabling the creation of new styles that are not present in the original dataset without requiring additional training of the prior. As shown in \Cref{fig:compositionalprior}, two different styles can be composed by blending their style-conditioned predictions from the diffusion model to produce new styles, such as ``\emph{AeroPlane + HighKnees}''. The composite prior is constructed via:
\begin{equation*}
    f_{\mathrm{comp}}
= M_{\mathrm{upper}} \odot f(\rvx^i, c_{\mathrm{aeroplane}})
+ M_{\mathrm{lower}} \odot f(\rvx^i, c_{\mathrm{highknees}}),
\end{equation*}
where $M_{\mathrm{upper}}$ and $M_{\mathrm{lower}}$ are binary masks applied to the upper- and lower-body features, respectively.  
The resulting prior $f_{\mathrm{comp}}$ can be used directly as the SMP reward $r^{\mathrm{smp}}$. Moreover, our proposed generative state initialization (GSI) can also adopt the composite prior to generate initial states of the new style, enabling effective exploration when training policies for the new style. Using SMP and GSI together with $f_{\mathrm{comp}}$, our framework is able to train an agent that follows task commands while spreading its arms, according to the ``AeroPlane'' style, and lifting its knees high, according to the ``HighKnees'' style, all without relying on any reference motion data of the composite style. We further demonstrate the flexibility of SMP by crafting more expressive and playful styles. For instance, we create an ``Elated + FlickLegs'' prior that combines the arm-swinging motion of the ``Elated'' style with exaggerated lateral leg flicks during locomotion, resulting in an energetic and humorous movement pattern. We also compose the ``GracefulArms'' and ``Spin'' styles to produce a ballroom-dance-like behavior, characterized by elegant arm motions and continuous rotational movements.

\subsection{One Motion Prior for Multiple Tasks}
\label{subsec:lafan}
To further demonstrate the \emph{reusability} of SMP, we show that a pretrained score-matching motion prior can be reused to train multiple policies for different tasks, such as steering, target location and dodgeball, as show in \Cref{tab:loco_task_perf}. The score-matching motion prior is trained on a subset of the LaFAN1 dataset \citep{harvey2020robust}, containing unstructured running behaviors. The resulting policies achieve high task returns while exhibiting naturalistic gaits. As shown in \Cref{fig:heading,fig:lafan_loc}, the character can dynamically transition between different gaits according to changing goals. The \emph{Steering} policy automatically executes a backward jog when the heading direction is oriented in an opposite direction with respect to the target direction, then the policy transitions to a forward jog once the two directions are aligned. The policy also learns lateral gaits, such as side-stepping and cross-stepping, when the heading direction is approximately orthogonal to the target direction. 

Policies trained with SMP also exhibit human-like strategies in the \emph{Target Location} task. When the target is far away, the character executes a fast running behavior. When the target is near, the character automatically slows down to slower walking gaits.  These nuanced behaviors arise purely from the score-matching motion prior combined with a simple target-distance task objective. No motion planner is required to explicitly specify which gait the character should perform in different scenarios. 

\begin{figure}[t]
	\centering
        \captionsetup{skip=0pt}
    \subfigure{\includegraphics[height=0.34\columnwidth]{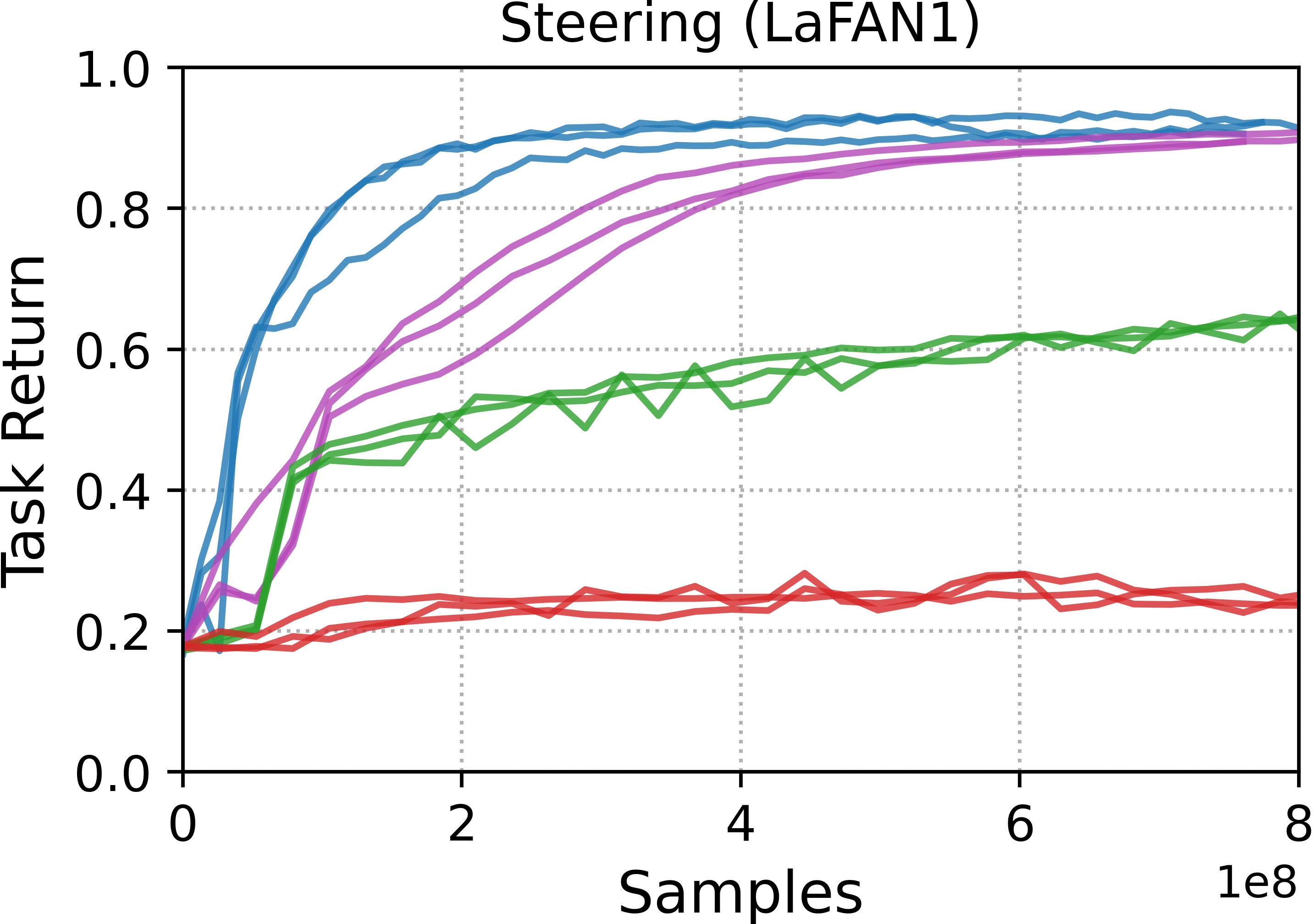}}
    \subfigure{\includegraphics[height=0.34\columnwidth]{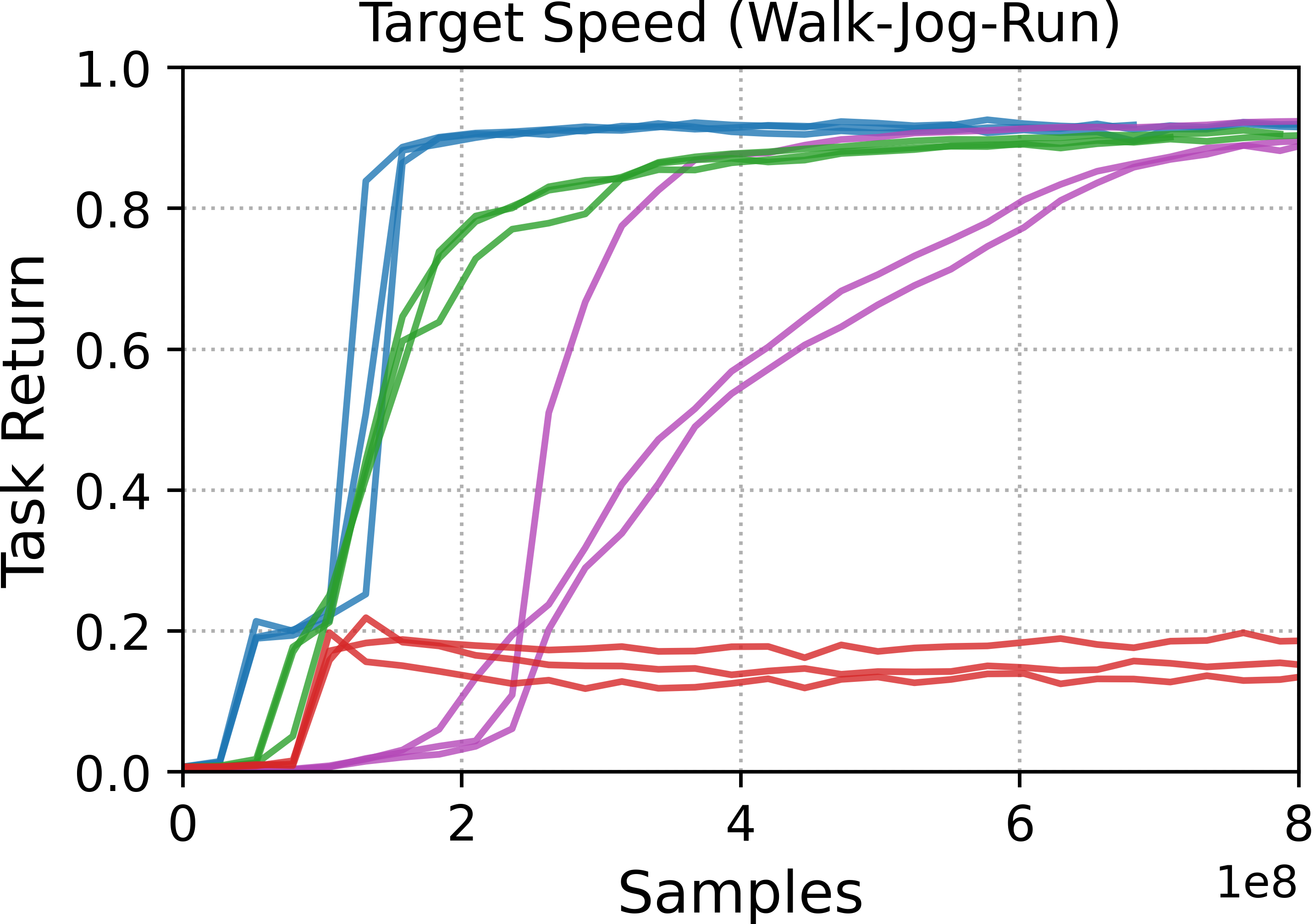}}\\ 
    \vspace{-6pt}
    \subfigure{\includegraphics[height=0.34\columnwidth]{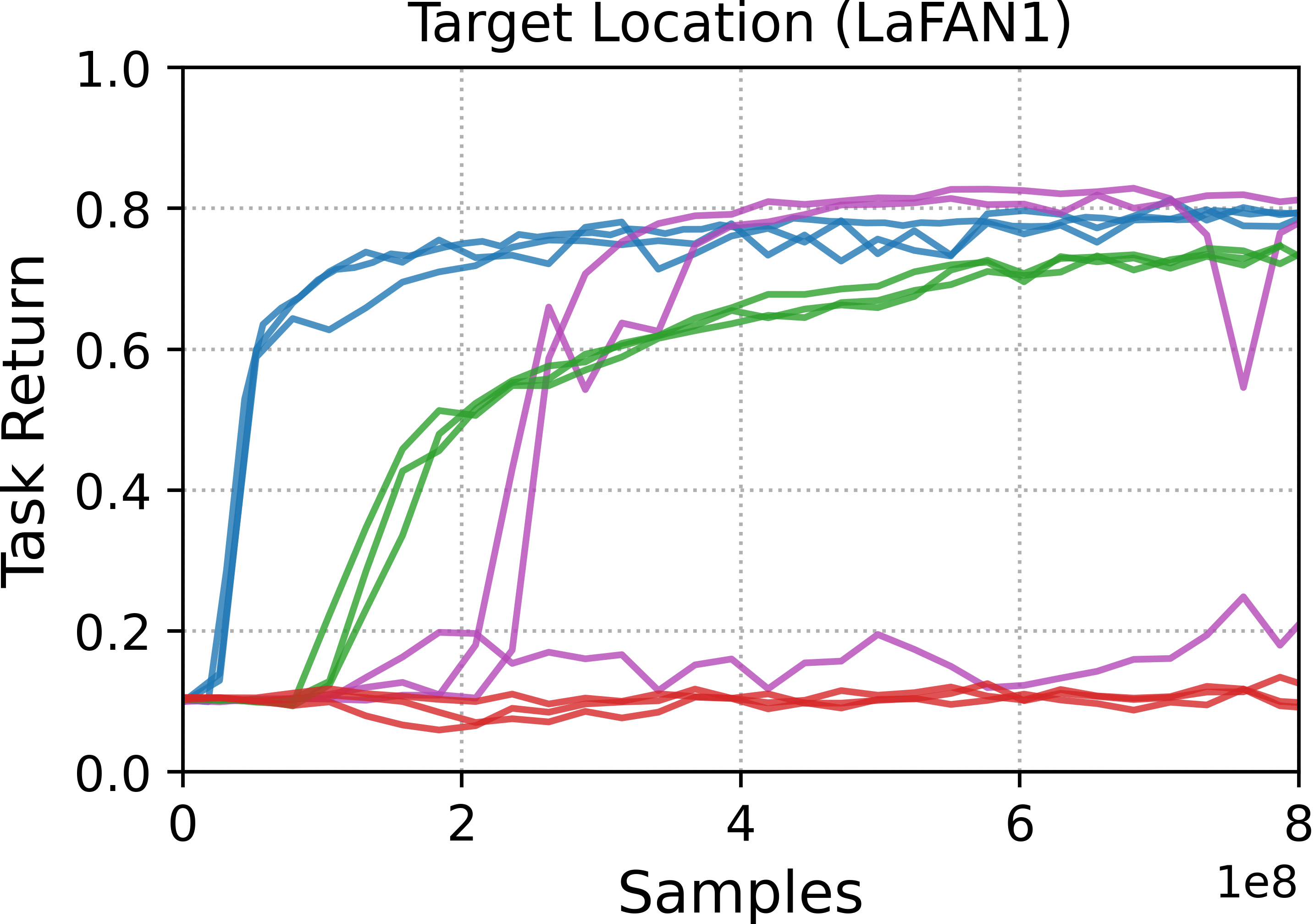}}
    \subfigure{\includegraphics[height=0.34\columnwidth]{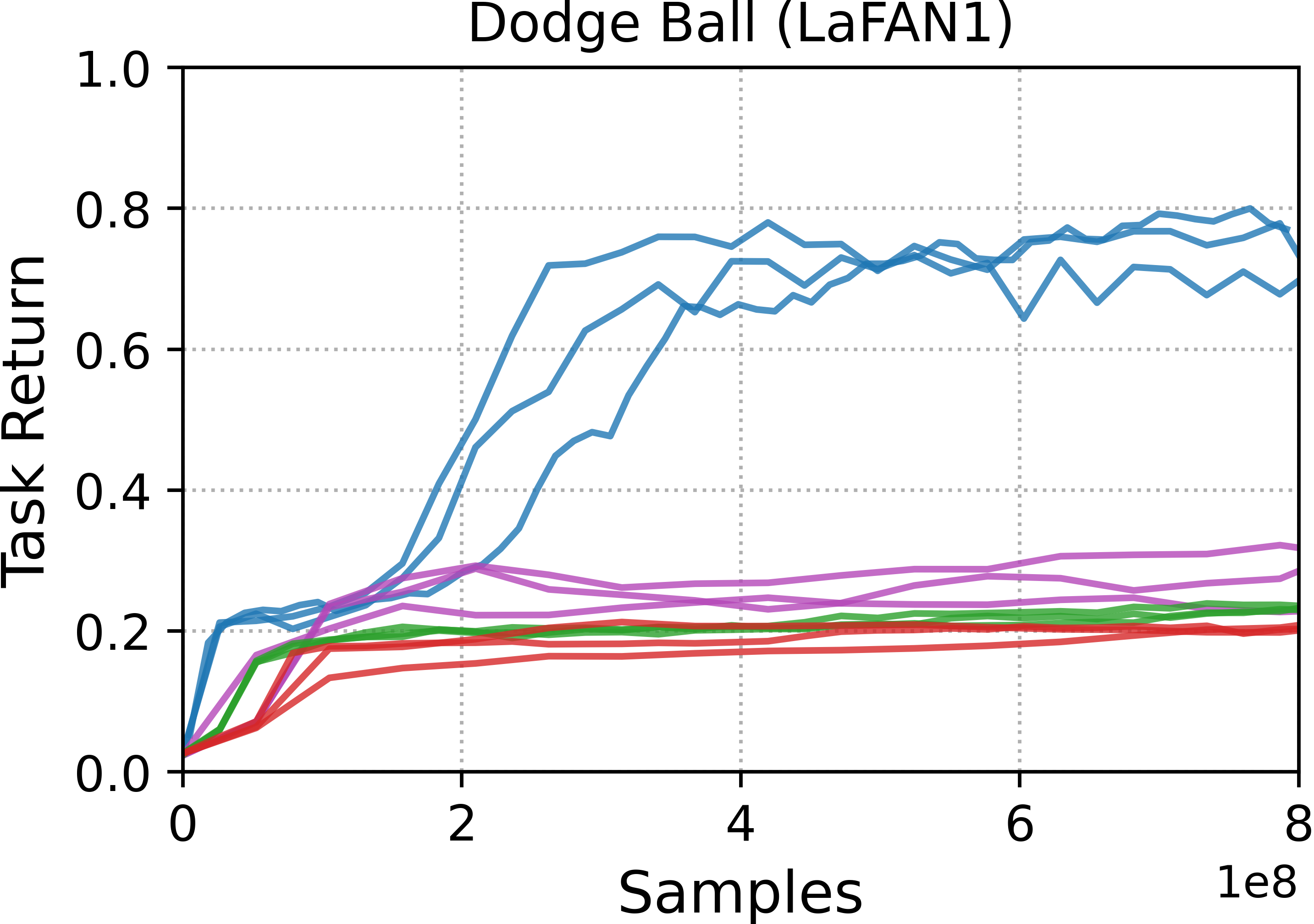}}\\
    \vspace{-0.2cm}
    \subfigure{\includegraphics[width=0.8\linewidth]{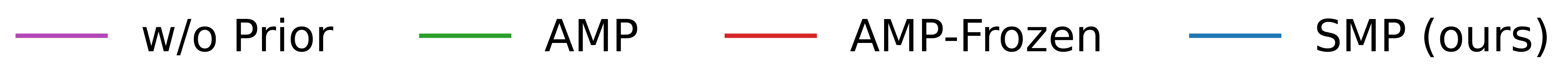}}\\
    \vspace{-0.1cm}
\caption{Comparison of normalized task returns across motion control tasks. SMPs demonstrate better sample efficiency, potentially due to the more stable and consistent RL provided by the stationary SMP reward function.}
\vspace{-0.2cm}
\label{fig:task_return_curves}
\end{figure}

SMP provides the policy with the flexibility to adapt behaviors in the dataset in order to create new skills for different tasks. In the challenging \emph{Dodgeball} task, agents trained with the locomotion prior spontaneously develop agile jumping and dodging skills, as shown in \Cref{fig:dodge}. These behaviors were not present in the original dataset, but they closely resemble real human strategies in dodgeball.
In contrast, AMP is unable to learn effective policies for this task, potentially due to the instability of the adversarial objective when the required dodging skill deviates significantly from the reference dataset, which contains only locomotion motions. Furthermore, policies trained using AMP with a frozen discriminator are unable to produce natural behaviors or achieve strong task performance across all tasks. This is because freezing and reusing the AMP discriminator to train new policies introduces severe out-of-distribution inputs for the discriminator. These out-of-distribution inputs can lead to incorrect reward values that cause the policy to adopt behaviors that are ill-suited for the task. This suggests that adversarial priors cannot be effectively reused, even for identical tasks. In comparison, SMP allows a single pretrained motion prior to be reused across different policies and tasks. Moreover, as shown by the task return curves in \Cref{fig:task_return_curves}, SMP tends to exhibit more sample-efficient training. This efficiency improvement may in part be attributed to the fixed reward model, which provides more stationary and stable guidance throughout the policy training process compared to the nonstationary reward model from AMP. 

\begin{table}[t]
\caption{
Performance of combining different motion priors with task objectives, as well as optimizing task objectives alone. Task returns are normalized to $[0,1]$. A single score-matching motion prior can be effectively reused to train policies across different tasks. Even with a small dataset of only three reference snippets, SMP still enables the agent to perform the target-speed task effectively and with natural motion.
}
\centering
\small
\setlength{\tabcolsep}{2pt}
\begin{tabular}{c|c|cccc}
\toprule
\multirow{3}{*}{\textbf{Dataset}} & \multirow{3}{*}{\textbf{Task}} & \multicolumn{4}{c}{\textbf{Task Return}} \\
 & & w/o Prior & AMP & \begin{tabular}{c} {AMP} \\[-0.5ex] {Frozen} \end{tabular} & \begin{tabular}{c} {SMP} \\[-0.5ex] {(Ours)} \end{tabular} \\
\midrule
\multirow{3}{*}{LaFAN1}  & Steering    & $0.901^{\pm 0.006}$ & $0.634^{\pm 0.019}$ & $0.243^{\pm 0.008}$ & \cellcolor{green!10}$0.914^{\pm 0.006}$\\
                        & Target Location   & $0.615^{\pm 0.268}$ & $0.737^{\pm 0.014}$ & $0.101^{\pm 0.012}$ & \cellcolor{green!10}$0.793 ^{\pm 0.006}$ \\
                        & Dodgeball   & $0.277^{\pm 0.046}$ & $0.233^{\pm 0.003}$ & $0.204^{\pm 0.004}$ & \cellcolor{green!10}$0.733 ^{\pm 0.035}$ \\
\midrule
\begin{tabular}{c} {Walk-Jog} \\[-0.5ex] {-Run} \end{tabular}            & Target Speed      & $0.905^{\pm 0.013}$ & $0.904^{\pm 0.002}$ & $0.158^{\pm 0.021}$ & \cellcolor{green!10}$0.918 ^{\pm 0.002}$ \\
\bottomrule
\end{tabular}
\label{tab:loco_task_perf}
\end{table}
\begin{table}[t]
\caption{
Performance of SMP on object carry and getup tasks. SMP can be extended to model human–object interaction priors and train effective interaction policies. SMP can also be used to train robust getup policies that can recover from arbitrary fallen states.}
\centering
\begin{tabular}{c|cc}
\toprule
{\textbf{Task}} & \textbf{Task Return} & \textbf{Success Rate}\\
\midrule
Object Carry& $0.909^{\pm 0.022}$ & $0.997^{\pm 0.004}$\\
\midrule
Getup & $0.897^{\pm 0.029}$ & $0.998^{\pm 0.003}$ \\
\bottomrule
\end{tabular}
\label{tab:additional_task_perf}
\end{table}

\subsection{Human-Object Interaction Priors}
\label{subsec:hoi}
To demonstrate the effectiveness of our system beyond locomotion tasks, we apply SMP to create \emph{interaction priors} that jointly capture both character and object motions. 
Given a human-object interaction (HOI) dataset, we train an unconditional HOI diffusion model $f(\rvx^i_\mathrm{char}, \rvx^i_\mathrm{obj})$, which jointly models the distribution of character motions $\rvx_\mathrm{char}$ and object motions $\rvx_\mathrm{obj}$. 
The object's motion is represented by its rotation and position, both expressed in the character’s local coordinate frame. 
The prior reward $r^{\mathrm{prior}}_t$ is computed following the same procedure described in \Cref{alg:SMP} to evaluate the naturalness of character–object interactions. 
This object-interaction SMP is combined with task rewards to train policies to accomplish complex, multi-stage \emph{Object Carry} tasks, where the agent walks to a box, picks it up, and then carries it to an arbitrary target location. The learned policy is able to successfully perform this task using natural, and coordinated whole-body manipulation skills.

\begin{figure}[t]
\centering
\includegraphics[width=\linewidth]{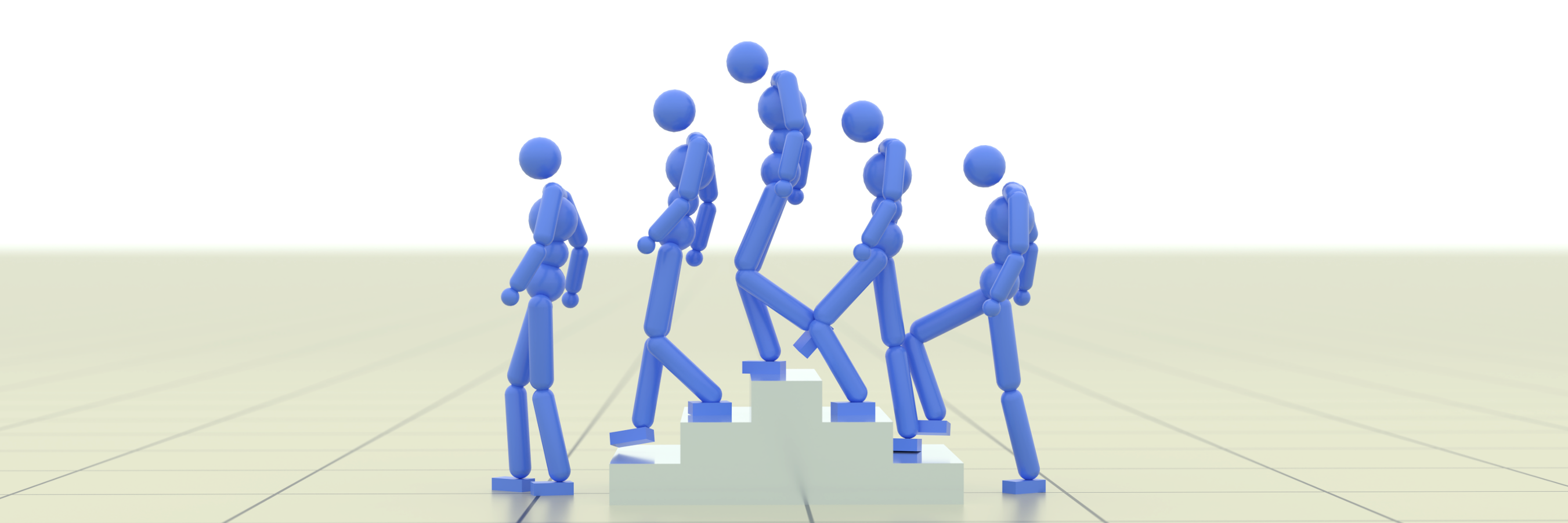}
\vspace{-0.5cm}
\caption{
Example of a stair traversal policy trained with an interaction prior. SMP enables the character to traverse multiple stairs with natural motions.
}
\label{fig:stairs}
\end{figure}

We also demonstrate that a \emph{Stair Traversal} policy can be trained using an interaction prior constructed from an unconditional human–scene interaction diffusion model. The position and orientation of the staircase are encoded in the character’s local coordinate frame and jointly modeled by the diffusion model. As shown in \Cref{fig:stairs}, the resulting policy enables the character to climb up and down the staircase with natural motions and stable foot contacts.

\subsection{Learning to Get Up}
\label{subsec:getup}
Getting up from arbitrary fallen states is an essential skill for both humans and humanoid agents. Naturalistic and physically plausible recovery behaviors are particularly important for safety and feasibility in real-world robotic systems~\citep{tao2022learning,he2025learning}. Producing such life-like get-up motions is challenging, as unconstrained policies often exploit highly dynamic or erratic movements that may succeed functionally in simulation but appear unnatural. As a result, many existing approaches rely on carefully designed auxiliary rewards or regularization terms to suppress these artifacts and encourage safe and smooth recovery behaviors~\citep{he2025learning}. Our results in \Cref{tab:additional_task_perf} demonstrate that SMP can successfully train a robust get-up policy capable of recovering from random fallen states by simply combining a body-height objective with a motion-prior objective, without requiring complicated hand-crafted regularization terms. In \Cref{fig:getup1}, after the character is knocked down, it naturally rolls over, pushes itself up with its hands, and returns to a standing posture. Additional qualitative results are provided in the supplementary video.

\subsection{Skill Emergence under Data Scarcity}
\label{subsec:wjr}

Reinforcement learning with distribution-matching objectives naturally allows agents to generalize and develop skill that not present in the dataset. 
In contrast to the typical application of SDS with image-based priors on large dataset, SMP can learn from as little as three seconds of motions, while still providing the agent with the flexibility to adapt and develop new behaviors beyond those in the original dataset. 
In the \emph{Target Speed} task, the training dataset contains only three motion clips at distinct speeds. Despite the limited reference motion data, the learned policy is able to adapt these motions to follow a continuous range of target speeds, as shown in \Cref{fig:speed}.
The character naturally adjusts its gait to match the target speed, walking at low target speeds, and transitioning into jogging and running as the speed increases. 
Although the reference dataset only contains motions at three discrete speeds, the policies learn to modulate the frequency and stride within each gait, producing continuous variations that preserve the style of the original motions while traveling at a broad range of speeds. 
Moreover, the policies exhibit smooth and naturalistic transitions between gaits that are not present in the dataset, including building up speed from a walk to a jog, decelerating from a run back to a walk, and bursting from a walk into a sprint.

As shown in \Cref{fig:task_return_curves}, the learning curves indicate that incorporating the motion prior significantly improves sample efficiency compared to the baseline trained without any prior reward (\emph{w/o Prior}). This improvement arises because the score-matching motion prior provides a dense and informative shaping signal that consistently biases exploration toward physically plausible and natural motions. By constraining the policy to remain within high-likelihood regions of the motion manifold during training, SMP reduces the need for trial-and-error exploration of unnatural behaviors, allowing the policy to focus its learning capacity on task-relevant control. As a result, task rewards and motion priors can reinforce each other, leading to faster convergence and more stable learning.

\subsection{Benchmark: Single-Clip Imitation}
\label{subsec:singleclip}
To further evaluate the effectiveness of SMP at imitating behaviors from reference motion clips, we compare our method with AMP \citep{peng2021amp}, AMP-Frozen, and SMILING \citep{wu2025diffusing} on a series of single-clip imitation tasks. All policies are trained without the task reward $r^g$. In addition to comparisons to prior distribution matching methods, we also include comparisons with a motion tracking method, DeepMimic (DM) \citep{peng2018deepmimic}. Imitation performance is assessed using the position tracking error $e_t^{\mathrm{POS}}$, as defined in \Cref{Eq:pos_tracking_err}.
Unlike motion-tracking methods such as DeepMimic, which are explicitly designed for precise replication of reference motions, SMP, AMP, and SMILING focus on imitating the general style of the motions. Therefore, these methods do not directly synchronize the policy with the reference trajectory. To ensure a more fair comparison across all methods, we follow the evaluation procedure of \citet{peng2021amp} and apply dynamic time warping (DTW) to SMP, AMP, and SMILING to temporally align the motions from the simulated character before calculating the position tracking error \citep{dtw_sakoe}.
Furthermore, pose-error termination is disabled for DeepMimic to ensure a consistent evaluation setup, as this early termination strategy is not applicable to non-tracking methods.

\begin{figure}[t]
	\centering
        \captionsetup{skip=0pt}
    \subfigure{\includegraphics[height=0.34\columnwidth]{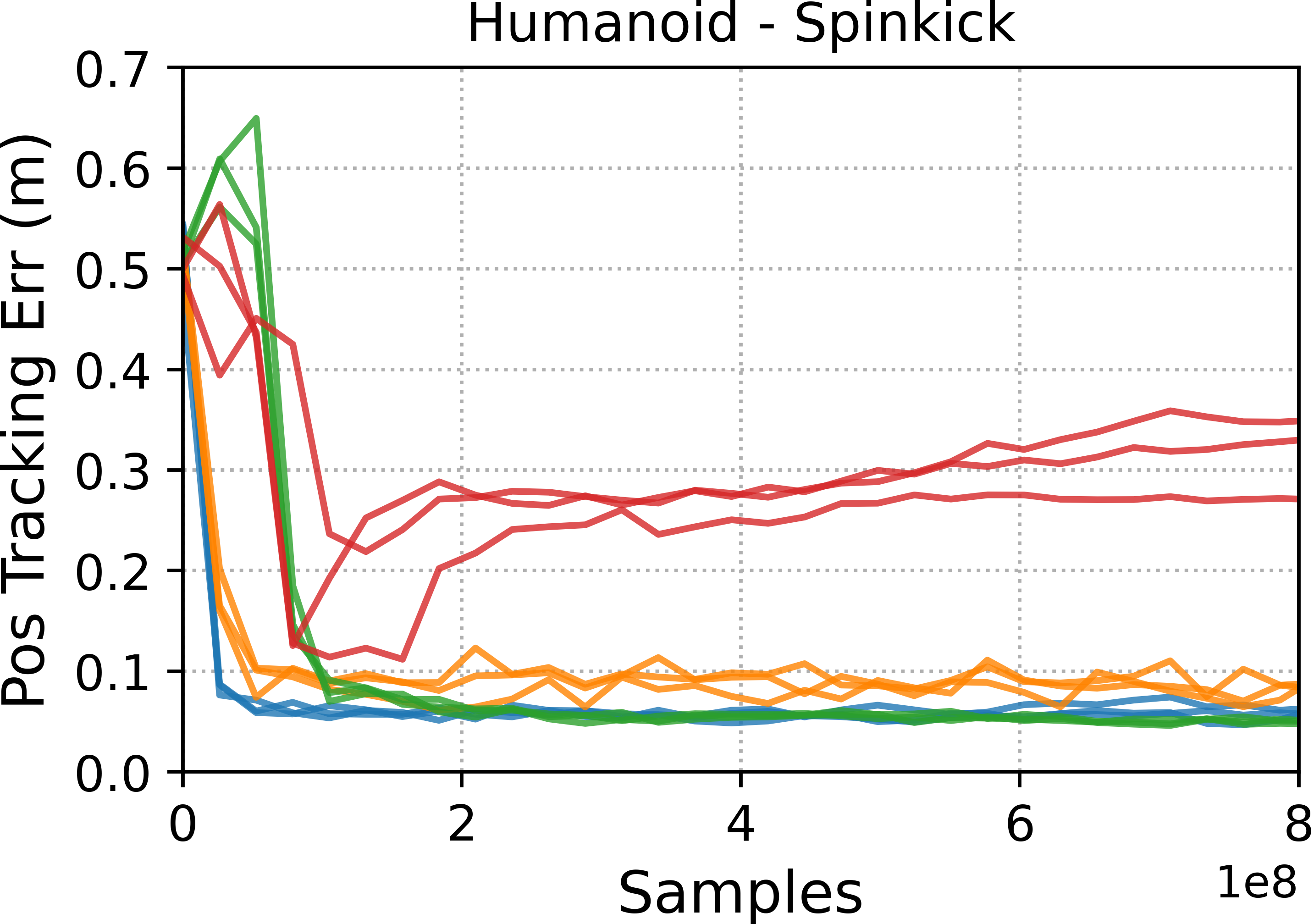}}
    \subfigure{\includegraphics[height=0.34\columnwidth]{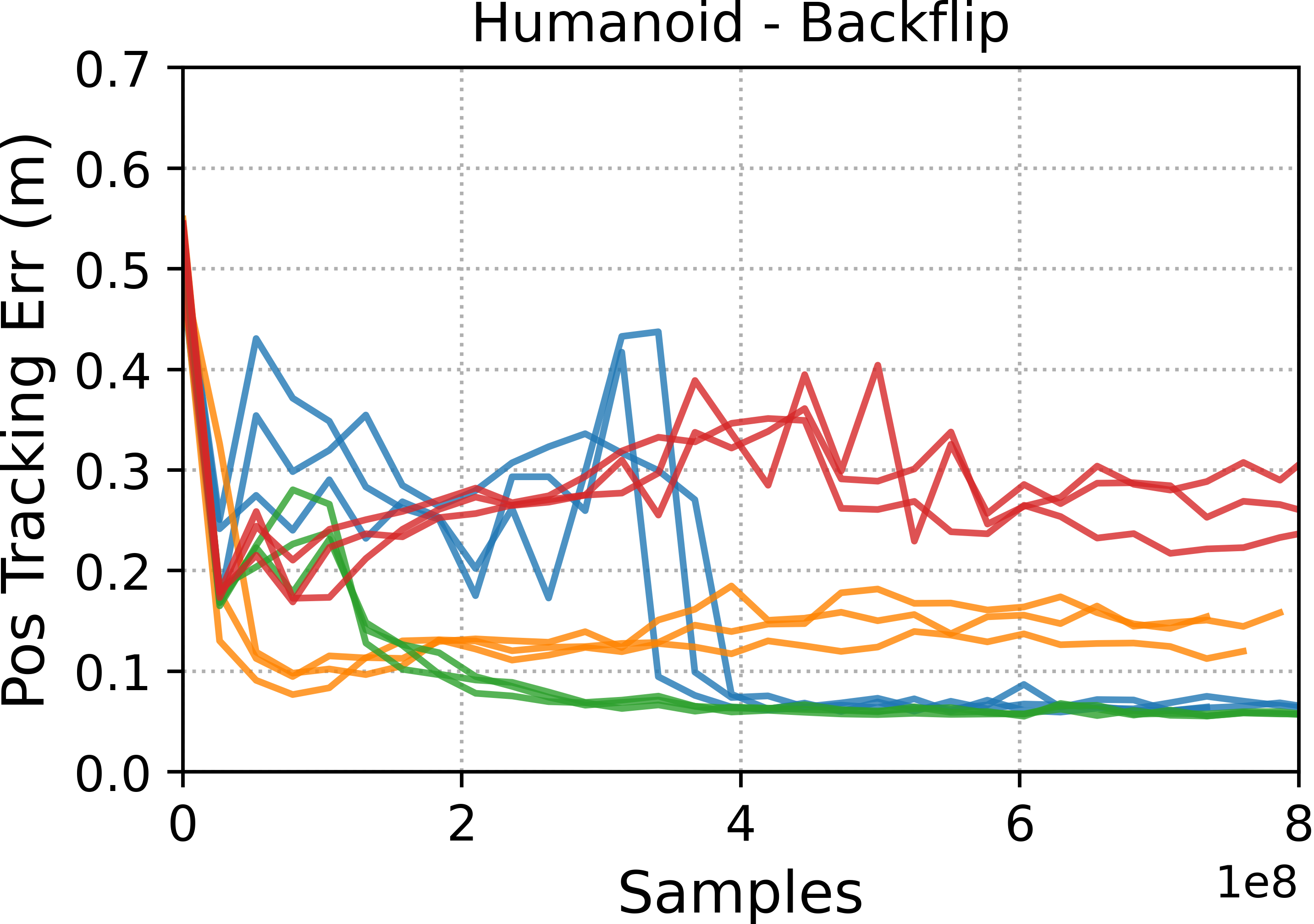}}\\ 
    \vspace{-6pt}
    \subfigure{\includegraphics[height=0.34\columnwidth]{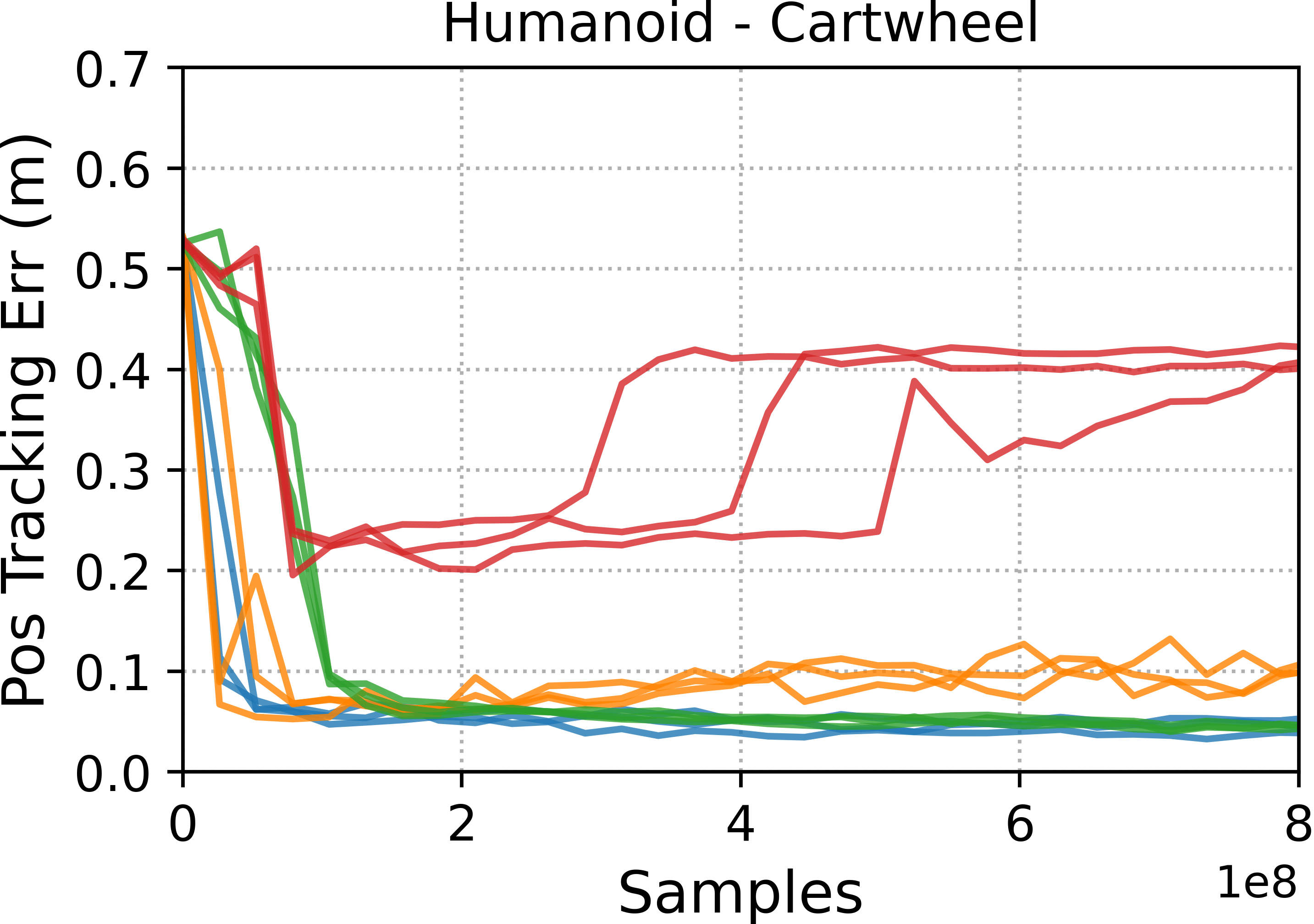}}
    \subfigure{\includegraphics[height=0.34\columnwidth]{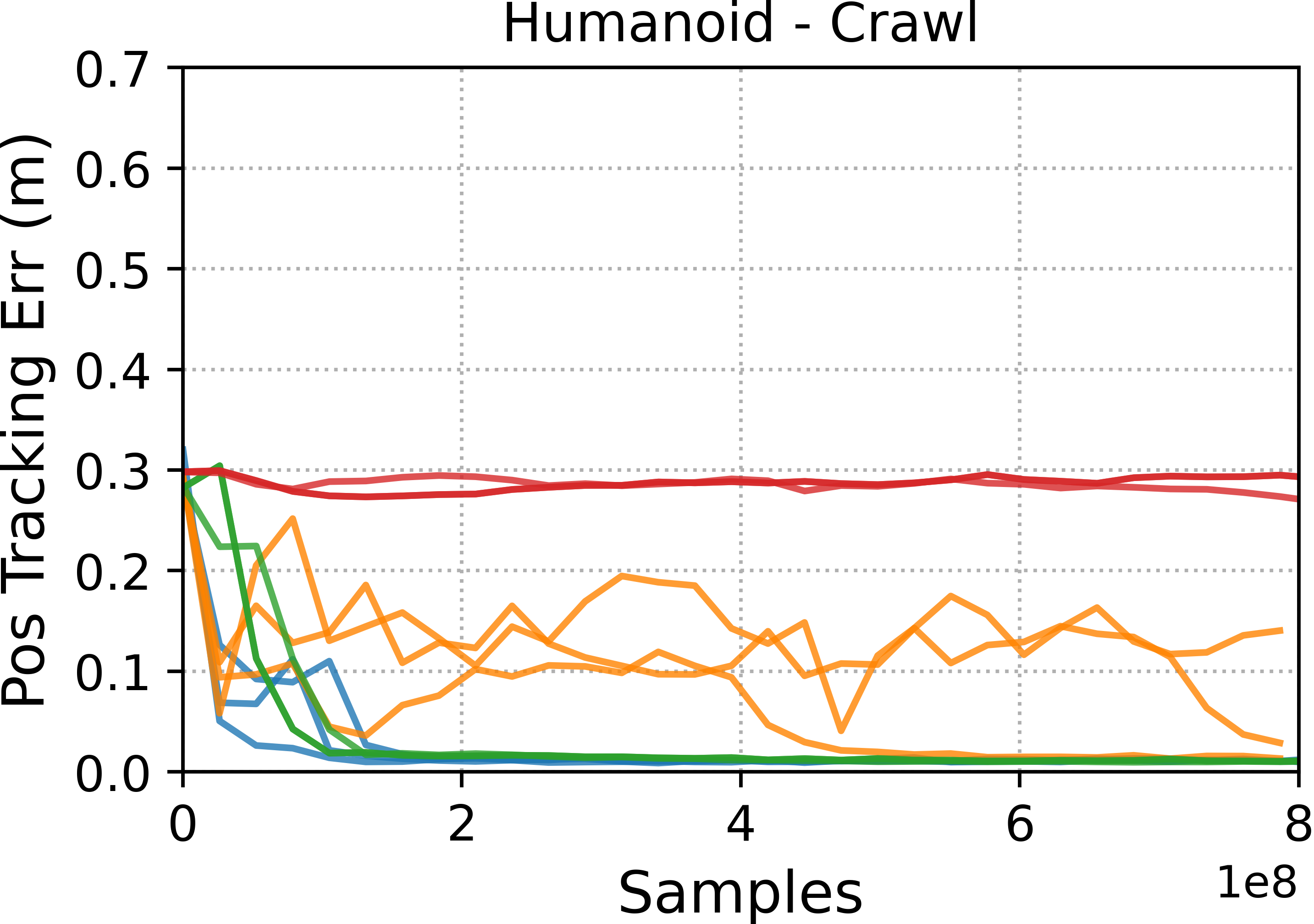}}\\
    \vspace{-6pt}
    \subfigure{\includegraphics[height=0.34\columnwidth]{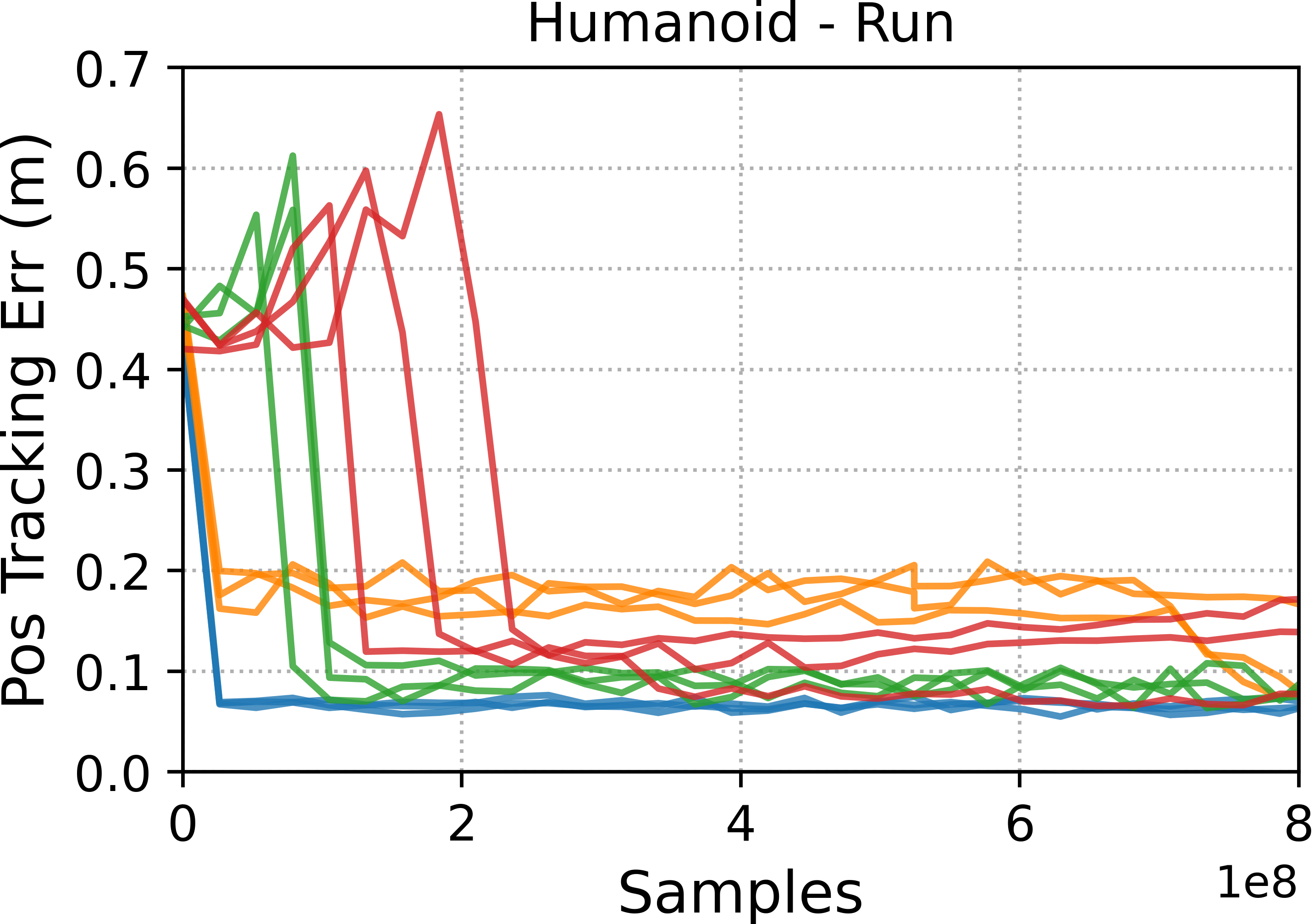}}
    \subfigure{\includegraphics[height=0.34\columnwidth]{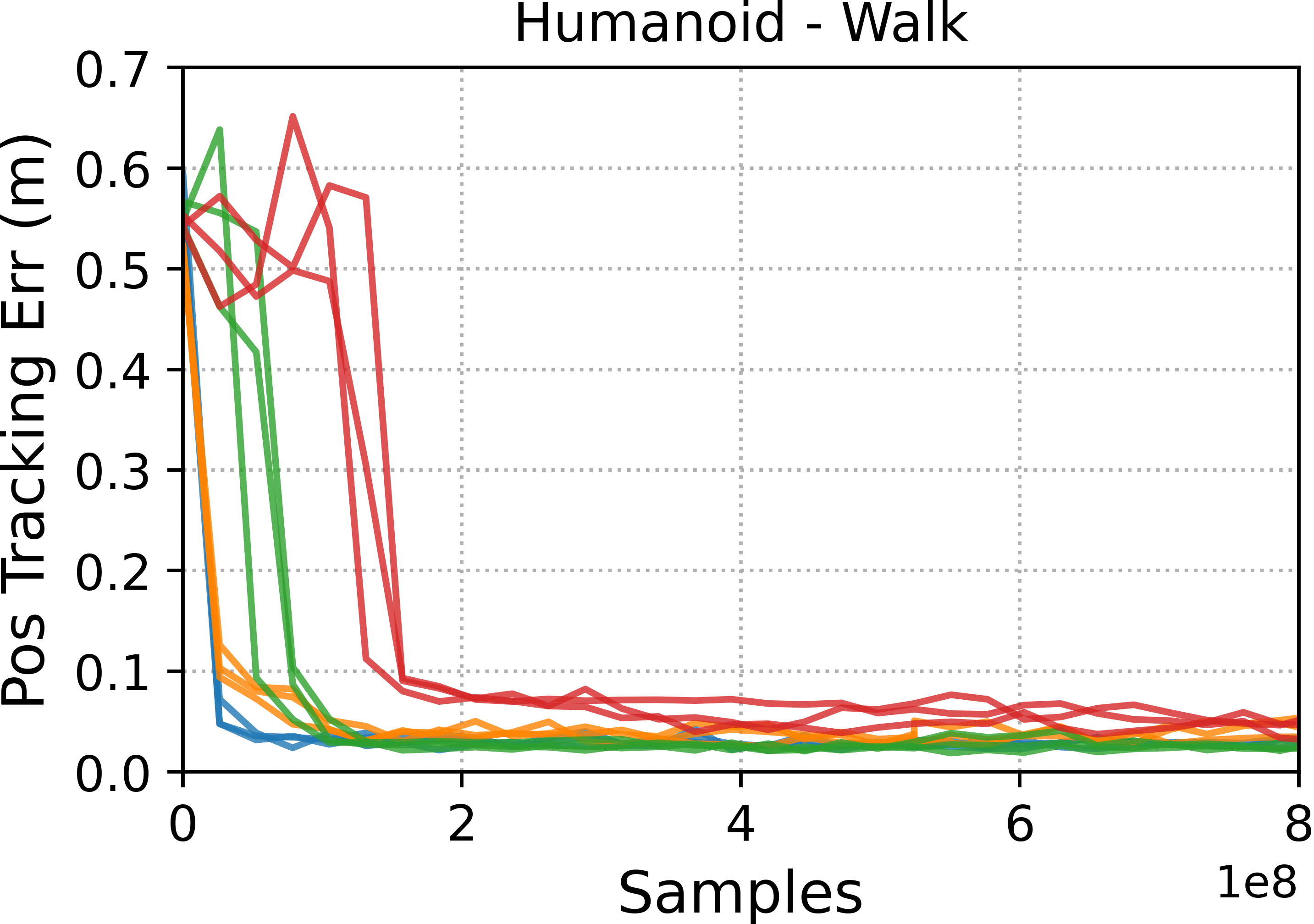}}\\
    \vspace{-0.2cm}
    \subfigure{\includegraphics[width=0.8\linewidth]{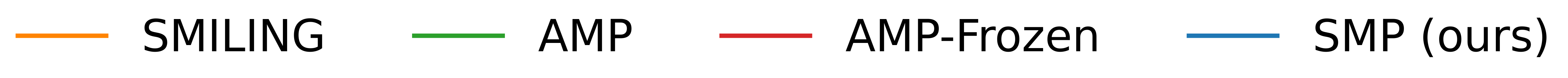}}\\
    \vspace{-0.1cm}
\caption{Learning curves of different methods on the single-clip imitation tasks. Three models initialized with different random seeds are trained for each method.
To evaluate the robustness of SMP, different diffusion models trained with different random seeds are used to train each SMP policy. SMP demonstrates consistent and effective performance across seeds.}
\label{fig:motion_imitation_learning_curve_selected}
\end{figure}

Quantitative results are summarized in Table~\ref{tab:single_clip}, with learning curves provided in \Cref{fig:motion_imitation_learning_curve_selected}. While AMP exhibits strong imitation performance, it requires retaining the reference motion data throughout policy training. In comparison, SMP accurately reproduces a diverse range of skills, including highly dynamic behaviors, such as backflip, and exhibits comparable imitation accuracy and sample efficiency to AMP, all without requiring access to the reference data during training.
Across skills, SMP consistently outperforms SMILING and AMP-Frozen.
AMP-Frozen, which attempts to remove AMP’s data dependency by simply using a fixed pre-trained discriminator, is unable to accurately reproduce the desired motions.
The tracking-based DeepMimic benefits from explicit synchronization between the reference motion and the policy, and thereby attains lower tracking errors on certain motions. However, SMP still performs competitively to this tracking-based method. DeepMimic's performance degrades on some challenging motions such as spinkick and cartwheel, where disabling pose-error termination leads to some training runs converging to suboptimal behaviors.

\begin{figure}[t]
	\centering
    \subfigure[Spinkick] {\includegraphics[width=1.0\columnwidth]{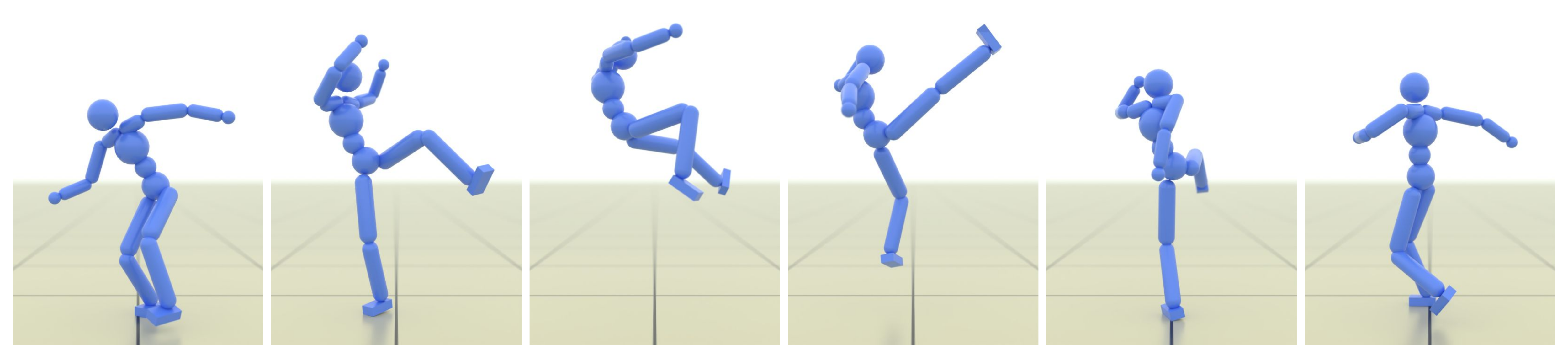}}\\
    \vspace{-0.4cm}
    \subfigure[Backflip]{\includegraphics[width=1.0\columnwidth]{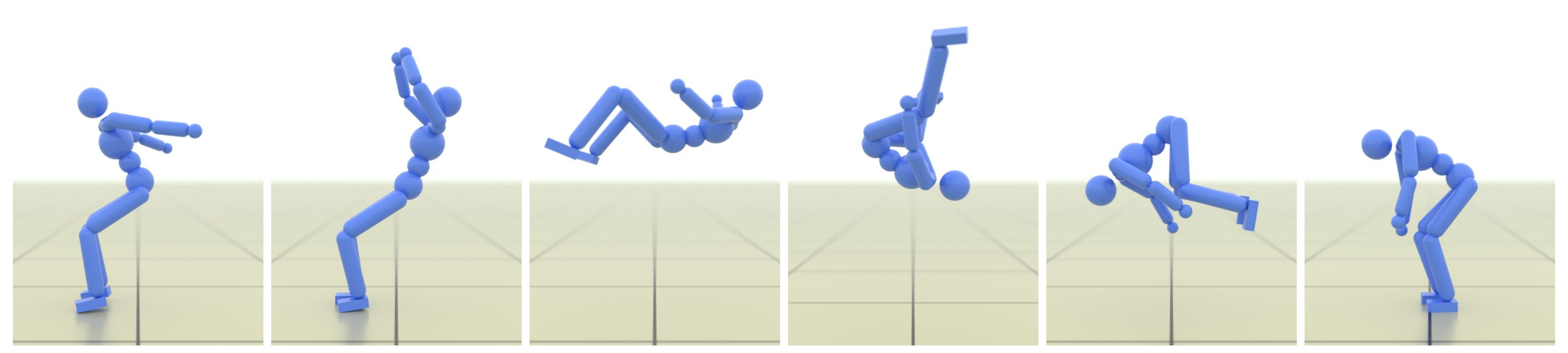}}\\
    \vspace{-0.4cm}
    \subfigure[Cartwheel]{\includegraphics[width=1.0\columnwidth]{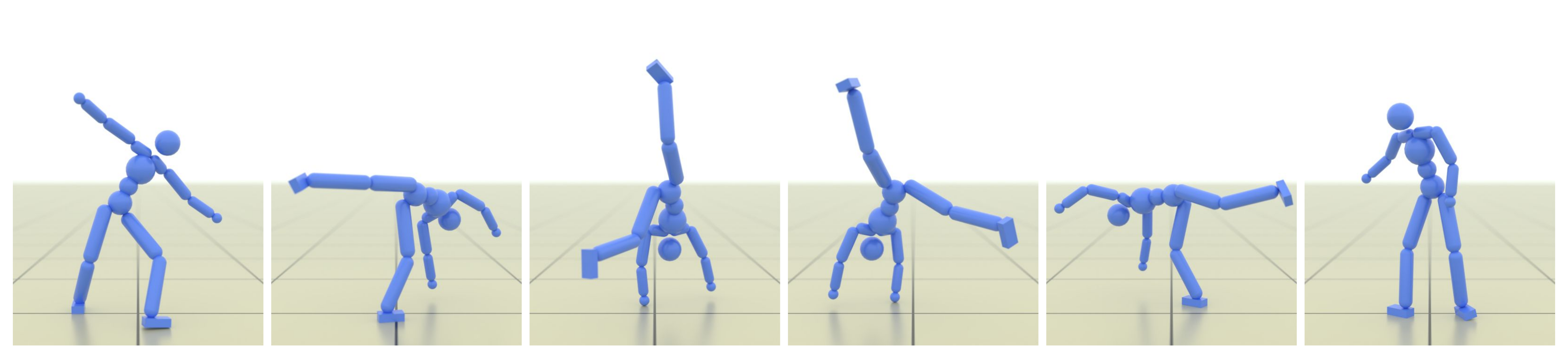}}\\
    \vspace{-0.4cm}
    \subfigure[Crawl]{\includegraphics[width=1.0\columnwidth]{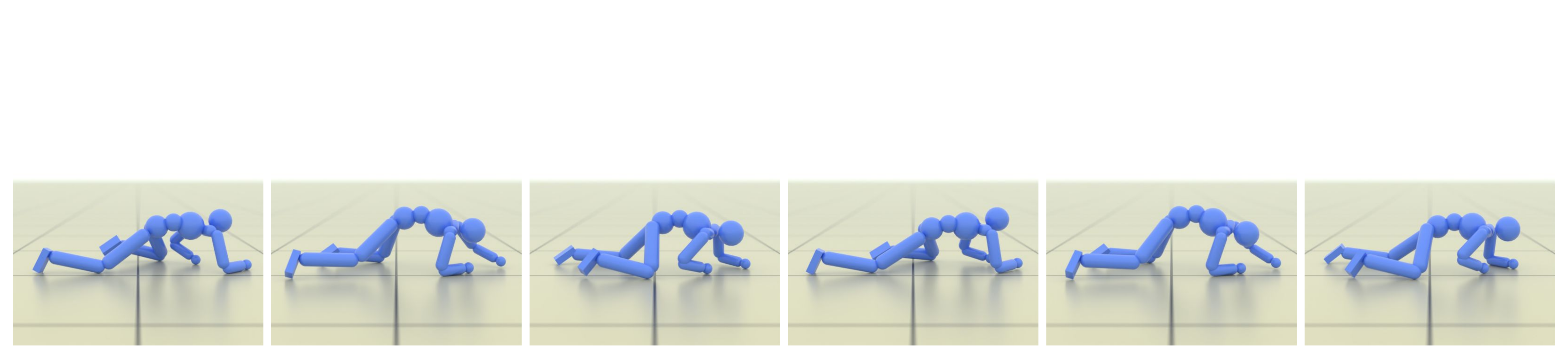}}\\
    \vspace{-0.4cm}
    \caption{Visual snapshots of humanoid characters trained via SMP imitating diverse skills, including highly dynamic and contact-rich motions.
    }
    \label{fig:single_clips_snapshots}
\end{figure}

\section{Real-World Robotic Applications}
\begin{figure}[t]
    \centering
    \subfigure[Steering] {\includegraphics[width=1.0\columnwidth]{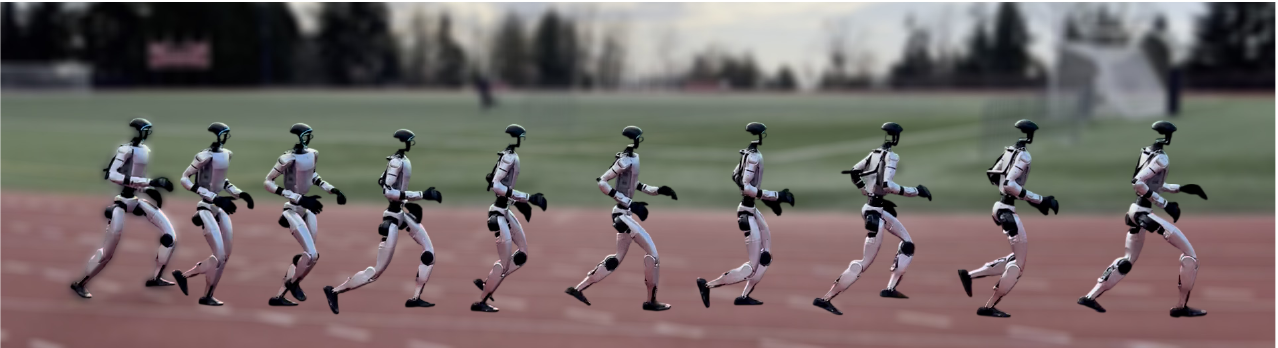}}\\
    \vspace{-0.3cm}
    \subfigure[Recovery Behaviors] {\includegraphics[width=1.0\columnwidth]{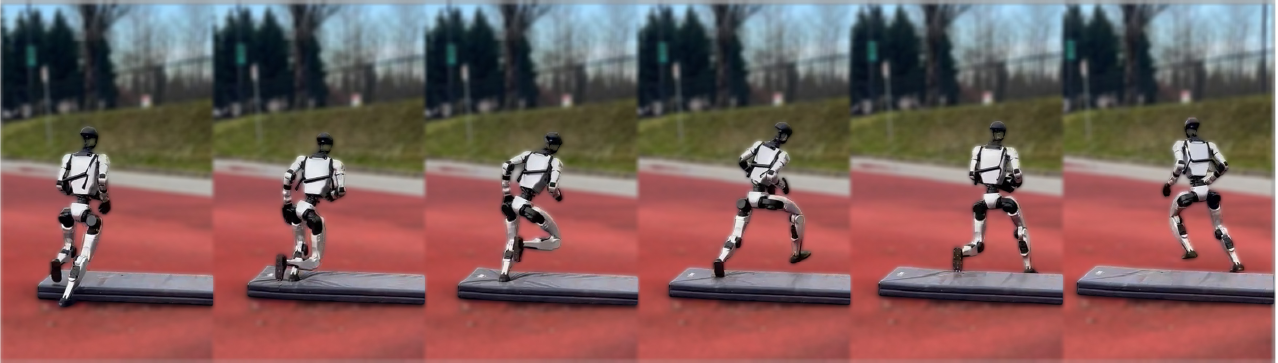}}\\
    \vspace{-0.3cm}
    \subfigure[Spinkick] {\includegraphics[width=1.0\columnwidth]{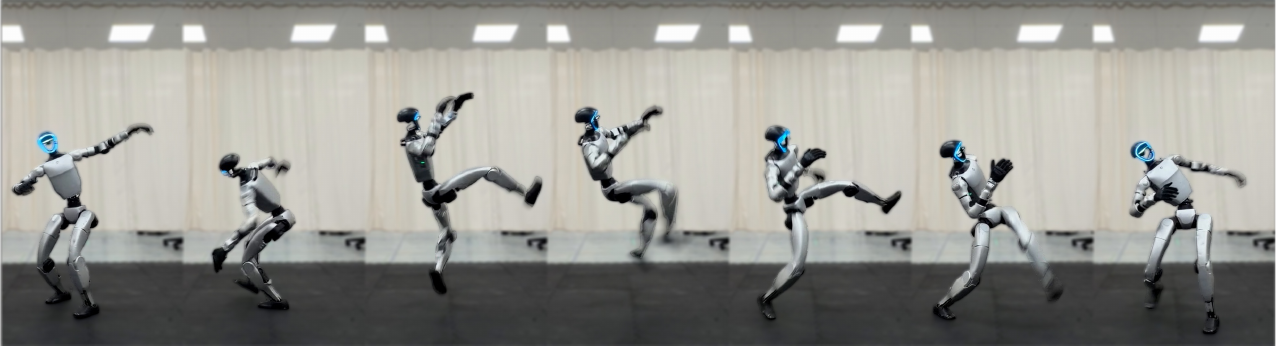}}\\
    \vspace{-0.3cm}
    \caption{SMP policies deployed on a Unitree G1 humanoid robot. The learned motor controllers can reproduce natural locomotion gaits, robust recovery behaviors, and agile motion skills in the real world.}
    \label{fig:robot_demo}
\end{figure}

In addition to experiments in simulated environments, SMP can be used to train controllers that enable real-world humanoid robots to perform natural, life-like behaviors. \Cref{fig:robot_demo} shows behaviors from a walking controller is trained entirely in simulation using SMP and then directly deployed on a G1 humanoid robot. 
Unlike the simulation experiments, the real-world controller has several modified in order to operate effectively in the real world. First, some state information available in simulation is difficult to reliably estimate on the physical robot. Therefore, the policy observes only a partial representation of the state consisting of proprioceptive information that can be recorded through onboard sensors:
\[
\mathbf{o}_t \triangleq \bigl[
\boldsymbol{\omega}_{t}^{\text{root}}\!,\,
\rvq_t^{\text{root}}\!,\,
\rvq_{t}\!,\,
\dot{\rvq}_t
\bigr] \in \mathbb{R}^{67},
\] 
which includes the root rotation $\rvq_t^{\text{root}} \in \mathbb{R}^{6}$, root angular velocity ${\boldsymbol{\omega}_{t}^{\text{root}} \in \mathbb{R}^3}$, local joint rotations $\rvq_t \in \mathbb{R}^{29}$, and local joint velocity $\dot{\rvq}_t \in \mathbb{R}^{29}$.

\begin{table}[t]
\caption{Position tracking error for individual skills. SMP successfully imitates a variety of skills, with imitation accuracy comparable to AMP while not requiring access to reference motion data during policy training. AMP-Frozen, which attempts to eliminate AMP’s data dependency by substituting a pre-trained discriminator, results in severely degraded behavior.}
\centering
\small
\setlength{\tabcolsep}{3pt}
\begin{tabular}{c|ccccc}
\toprule
\multirow{3}{*}{\textbf{Skill}} & \multicolumn{5}{c}{\textbf{Position Tracking Error [m]}} \\
& \multicolumn{1}{c|}{{DM}} & {AMP} & \begin{tabular}{c} {AMP} \\[-0.3ex] {Frozen} \end{tabular} & {SMILING} & {\begin{tabular}{c} {SMP} \\[-0.5ex] {(Ours)} \end{tabular}} \\
\midrule
Walk      & \multicolumn{1}{l|}{$0.010^{\pm 0.001}$} & $0.028 ^{\pm 0.004}$ & $0.044 ^{\pm 0.010}$ & $0.042^{\pm 0.007}
$ & $0.030^{\pm 0.004}$ \\

Run      & \multicolumn{1}{l|}{$0.013^{\pm 0.000}$} & $0.088 ^{\pm 0.010}$ & $0.129 ^{\pm 0.039}$ & $0.115^{\pm 0.040}$ & $0.067^{ \pm 0.001}$ \\

Spinkick     & \multicolumn{1}{l|}{$0.073^{\pm 0.061}$}  &$0.049 ^{\pm 0.001}$ & $0.324^{\pm 0.040}$ & $0.088^{\pm 0.005}$ & $0.059^{ \pm 0.006}$ \\

Cartwheel & \multicolumn{1}{l|}{$0.243^{\pm 0.157}$} &$0.043 ^{\pm 0.002}$ & $0.419^{\pm 0.013}$ & $0.104^{\pm 0.005}
$ & $0.043 ^{\pm 0.005}$ \\
Backflip      &  \multicolumn{1}{l|}{$0.073^{\pm 0.001}$} & $0.058^{\pm 0.002}$ & $0.272^{\pm 0.034}
$ & $0.144^{\pm 0.017}$ & $0.069^{\pm 0.008}$
 \\
Crawl &  \multicolumn{1}{l|}{$0.006^{\pm 0.000}$} & $0.011^{\pm 0.000}$ & $0.285^{\pm 0.008}$ & $0.061^{\pm 0.057}$ & $0.011^{\pm 0.001}$ \\
\midrule
Average & \multicolumn{1}{c|}{$0.070$} & \cellcolor{green!10}$0.046$ & $0.246$ & $0.092$ & \cellcolor{green!10}$0.046$ \\
\bottomrule
\end{tabular}
\label{tab:single_clip}
\end{table}

To ensure accurate reward and value estimation during training in simulation, we adopt an asymmetric actor-critic framework in which the SMP diffusion model and value function receive additional privileged information that are not provided to the policy \citep{asymetricActorCritic}. The motion features used by the diffusion model to compute the SMP reward $r^{smp}$ are identical to those described in \Cref{subsec:motion_rep}. The critic (\emph{i.e.}, value function) observes additional privileged state information, such as the root height, root linear velocity, and end-effector positions relative to the root. All features are recorded in the robot's local coordinate frame. To improve the robustness of the policy to variations in the system dynamics, domain randomization is applied during training in simulation \citep{DRPeng}, which randomized physical parameters, such as mass, friction and restitution. External perturbations are also applied during training to further enhance the policy's robustness.

In real-world experiments, the robot demonstrates robust locomotion under external perturbations, automatically adjusting its stride to maintain balance and continue moving forward. These recovery behaviors emerge automatically without relying on a runtime motion planner or manually designed heuristics. Additional qualitative results are provided in the supplementary video.

\section{Ablation and Sensitivity Analysis}
Our system is one of the first frameworks to construct a \emph{reusable} motion prior based on score-matching, that can achieve high fidelity results comparable to state-of-the-art adversarial imitation learning methods. In this section, we identify the key design decisions that enable more stable training and higher-quality results.

\paragraph{Ensemble Score-Matching}
We compare the proposed ensemble score-matching approach (``Random''), which averages multiple SDS evaluations over a fixed set of diffusion timesteps, against the typical strategy used in prior work, which evaluates SDS at a single randomly sampled timestep (``Random'')~\citep{poole2022dreamfusion, luo2024textaware, wu2025diffusing}. This comparison is conducted on the single-clip imitation tasks using reference motions of varying difficulties, including Run, Cartwheel, and Backflip.
As shown in \Cref{tab:fixed_set_sds}, the single-timestep SDS objective performs comparably on simple skills, but struggles on more challenging behaviors. For instance, when imitating a backflip, policies trained with a single SDS evaluation tend to collapse to standing still or falling backwards instead of executing a complete flip. In contrast, ensembling SDS evaluations across a fixed set of diffusion timesteps provides a more consistent and informative reward signal, which enables effective policy training even for challenging acrobatic motions.

\begin{table}[t]
\caption{
Comparison between computing the SDS objective using a single randomly sampled timestep versus ensemble over a fixed set of timesteps. 
Policies trained with random sampled timestep exhibit higher errors, particularly on challenging skills such as the backflip.
}
\centering
\small
\begin{tabular}{c|>{\centering\arraybackslash}p{1.8cm}
                    >{\centering\arraybackslash}p{1.8cm}}
\toprule
\multirow{2}{*}{\textbf{Skill}}& \multicolumn{2}{c}{\textbf{Position Tracking Error [m]}} \\
& Random & Ensemble \\
\midrule
Run & $0.062^{\pm 0.000}$ &$0.067 ^{\pm 0.001}$ \\
Cartwheel & $0.058^{\pm 0.015}$ & $0.043 ^{\pm 0.005}$ \\
Backflip & $0.195^{\pm 0.006}$ & $0.069 ^{\pm 0.008}$ \\
\midrule
Average &  0.105 & \cellcolor{green!10}0.060 \\
\bottomrule
\end{tabular}
\label{tab:fixed_set_sds}
\end{table}

\begin{table}[t]
\caption{
Effect of applying adaptive normalization (AdaNorm) to SDS errors. Position tracking errors are estimated with 3 runs initialized with different seeds for each motion, where each training run uses different independently trained diffusion models. AdaNorm improves robustness to variation across pretrained diffusion priors and motion skills.
}
\centering
\small
\begin{tabular}{c|>{\centering\arraybackslash}p{1.8cm}
                    >{\centering\arraybackslash}p{1.8cm}}
\toprule
\multirow{2}{*}{\textbf{Skill}}& \multicolumn{2}{c}{\textbf{Position Tracking Error [m]}} \\
& w/o AdaNorm & w/ AdaNorm \\
\midrule
SpinKick & $0.073^{\pm 0.002}$ &$0.059 ^{\pm 0.006}$ \\
Cartwheel & $0.177^{\pm 0.115}$ & $0.043 ^{\pm 0.005}$ \\
Backflip & $0.279^{\pm 0.110}$ & $0.069 ^{\pm 0.008}$ \\
\midrule
Average &  0.176 & \cellcolor{green!10}0.057 \\
\bottomrule
\end{tabular}
\label{tab:adanorm_ablation}
\end{table}

To further analyze the statistical effect of ESM, we evaluate the SDS error of the same motion, a HighKnees clip, 1024 times. With ESM, the variance drops to $9.964 \times 10^{-6}$, compared to $1.140$ without ESM, while the mean remains comparable at $1.339$ versus $1.309$. This confirms that ESM substantially reduces variance without introducing significant bias, thereby improving stability of RL training.

\paragraph{Adaptive Normalization of SDS Errors}
We evaluate the effect of adaptive normalization of the SDS error across different diffusion model checkpoints and motion skills. As shown in \Cref{tab:adanorm_ablation}, when using the same training configuration across multiple diffusion checkpoints, policies trained using SMP \emph{without} adaptive normalization exhibit larger performance variations, indicating greater sensitivity to hyperparameter settings and a greater need for careful manual tuning. In contrast, policies trained \emph{with} adaptive normalization achieve consistently strong performance and low position tracking errors, resulting in higher-quality motions while substantially reducing the burden of manual hyperparameter tuning.

\begin{table}[t]
\caption{
Comparison of using different diffusion timesteps $\sK$ for reward calculation. 
Ensemble score-matching (ESM) enables SMP to be robust to the choice of timesteps. Using diffusion timesteps $[22, 15, 8]$ consistently leads to better performance across tasks and is the default configuration used in all experiments unless stated otherwise.
}
\centering
\small
\setlength{\tabcolsep}{3pt}
\begin{tabular}{c|ccc}
\toprule
\multirow{2}{*}{\textbf{Skill}}& \multicolumn{3}{c}{\textbf{Position Tracking Error [m]}} \\
 & [15, 8, 1] & [43, 36, 29]& [22, 15, 8]\\
\midrule
Run  &$0.068 ^{\pm 0.006}$ &$0.059 ^{\pm 0.006}$& $0.067^{\pm 0.001}$\\
Cartwheel  &$0.055 ^{\pm 0.000}$ &$0.172 ^{\pm 0.052}$& $0.043^{\pm 0.005}$\\
Backflip  &$0.067 ^{\pm 0.009}$ &$0.064 ^{\pm 0.005}$& $0.069^{\pm 0.008}$\\
\midrule
Average  & 0.063 & 0.098 & \cellcolor{green!10}0.060\\
\bottomrule
\end{tabular}
\label{tab:diffusion_ts_ablation}
\end{table}

\begin{figure}[t]
\centering
    \captionsetup{skip=0pt}
    \subfigure{\includegraphics[height=0.34\columnwidth]{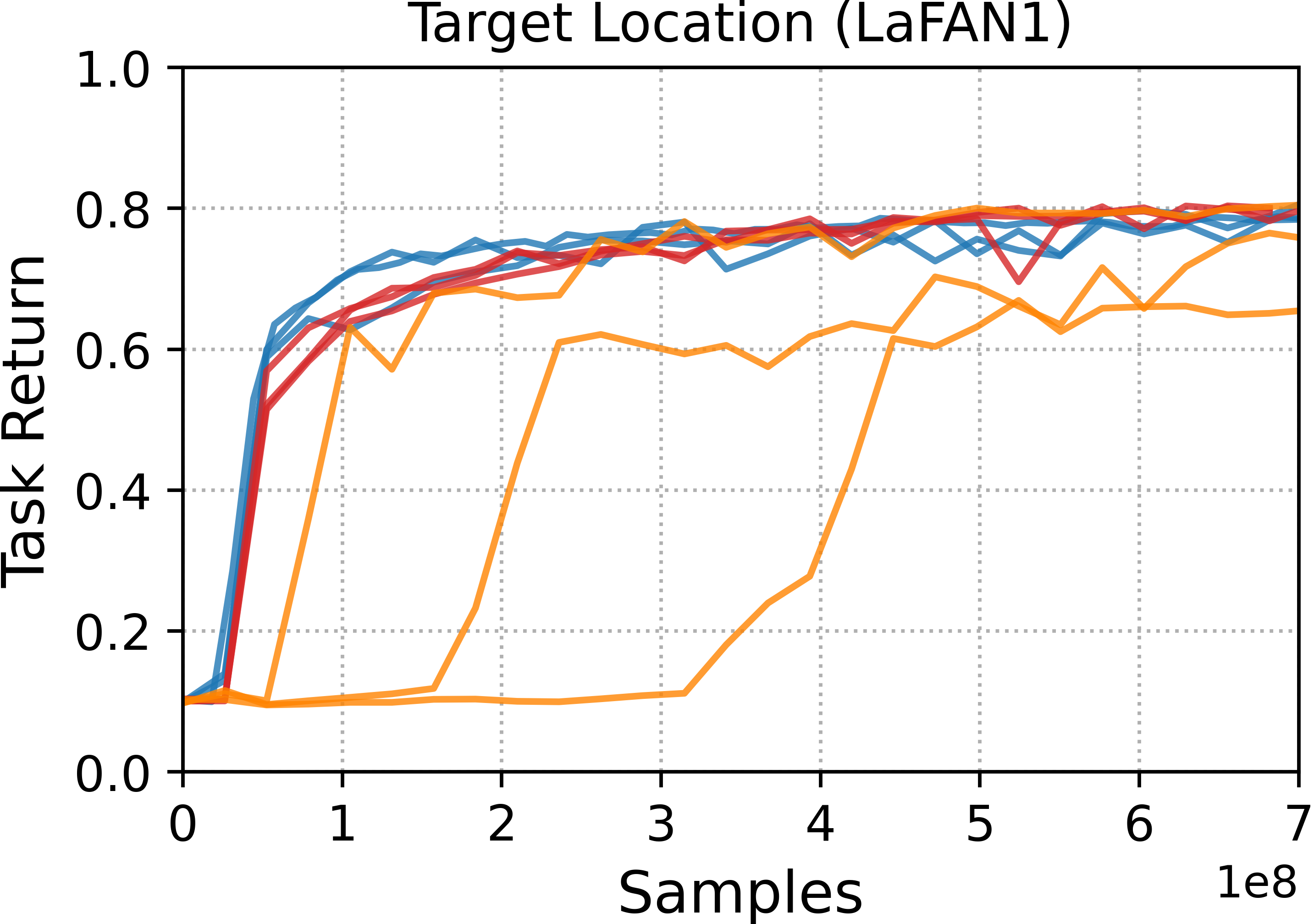}}
    \subfigure{\includegraphics[height=0.34\columnwidth]{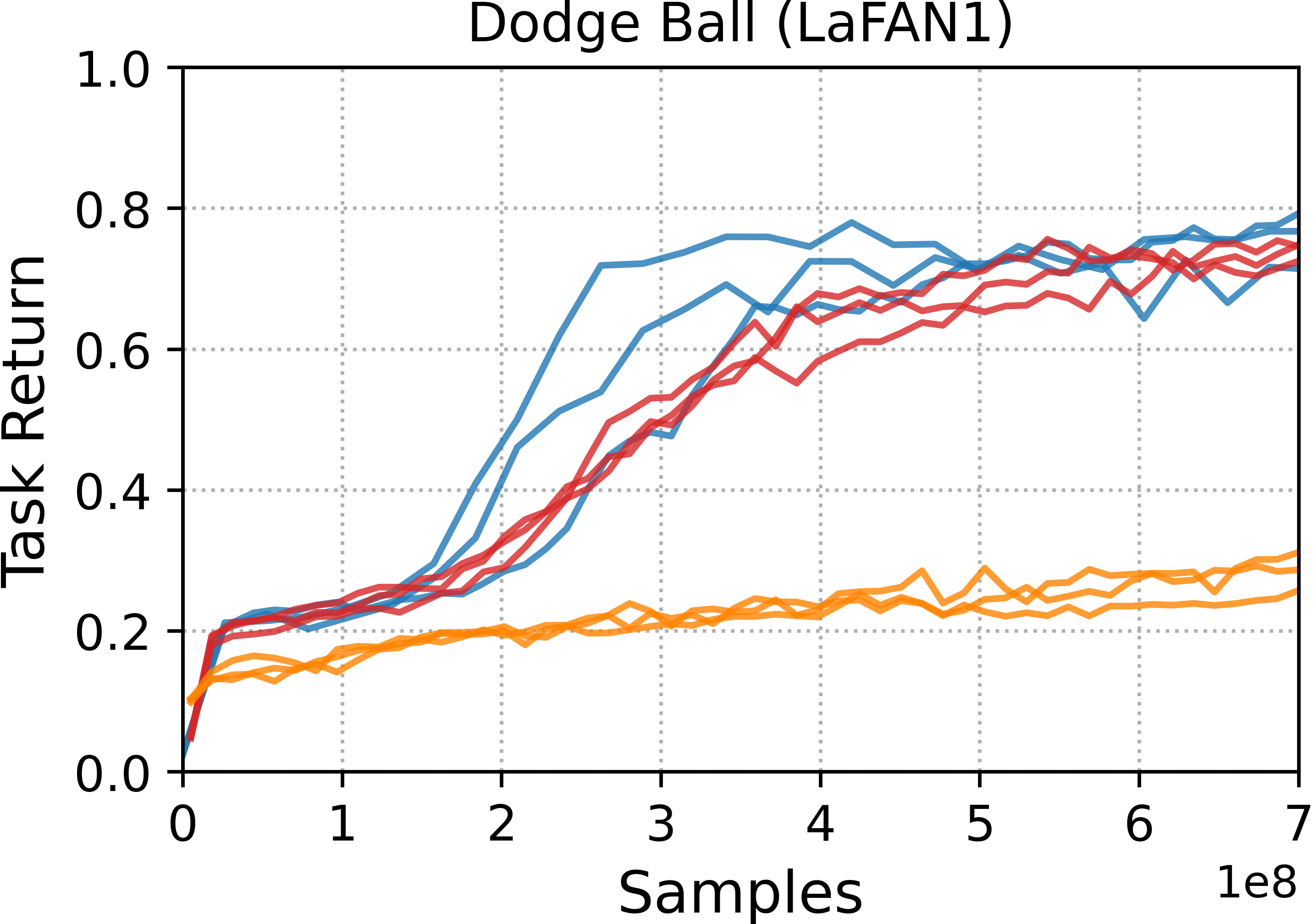}}\\ 
    \vspace{-2pt}
    \includegraphics[width=0.65\columnwidth]{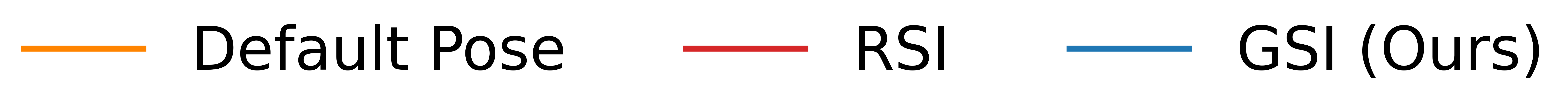}\\ 
\caption{Learning curves comparing the performance of policies trained with different state initialization strategies: default pose (T-pose) initialization, reference state initialization (RSI), and generative state initialization (GSI). GSI exhibits similar task performance and sample efficiency as RSI, without requiring initial states to be sampled from the original motion clips.}
\label{fig:state_init_ablation}
\end{figure}

\paragraph{Diffusion Timestep Choice}
We evaluate the effect of different diffusion timestep sets $\sK$ used in the SMP reward calculation (\Cref{eq:averageprior}) on single-clip imitation tasks. As shown in \Cref{tab:diffusion_ts_ablation}, different choices of diffusion timesteps yield similar performance, demonstrating that SMP is robust to timestep selection due to ESM. However, a trade-off remains between the reliability and precision of the SMP reward signal. Smaller diffusion timesteps provide more precise and fine-grained guidance, but are less reliable for out-of-distribution samples; this limitation is less critical for simpler tasks such as single-clip imitation, which do not require long-horizon exploration or generalization. In contrast, larger diffusion timesteps yield more reliable but less informative rewards, which can be insufficient for imitating challenging skills that require precise guidance. Across all evaluated tasks, we find that the timestep set $[22, 15, 8]$ consistently achieves the best performance, and we adopt this setting for all experiments unless otherwise stated. However, careful selection of the diffusion timesteps could further improve performance for specific tasks.

\paragraph{Generative State Initialization}
Effective state initialization strategies are crucial for improving exploration efficiency in imitation learning. We compare three initialization strategies for the target location task and the dodgeball task: default pose (T-pose) initialization, reference state initialization (RSI), and generative state initialization (GSI). Each control policy is trained under identical conditions, differing only in the initialization method. As shown in \Cref{fig:state_init_ablation}, RSI achieves better sample efficiency than T-pose initialization by initializing the agent in states closer to the reference motion distribution. Our proposed GSI achieves performance and sample efficiency comparable to RSI, while eliminating the need for access to the original motion dataset during policy training. This result highlights scenarios where reference motion data are unavailable or cannot be retained, and demonstrates that a fully \emph{modular} motion-prior reward framework is feasible. 

\begin{figure}[t]
\centering
    \captionsetup{skip=0pt}
    \includegraphics[width=\linewidth]{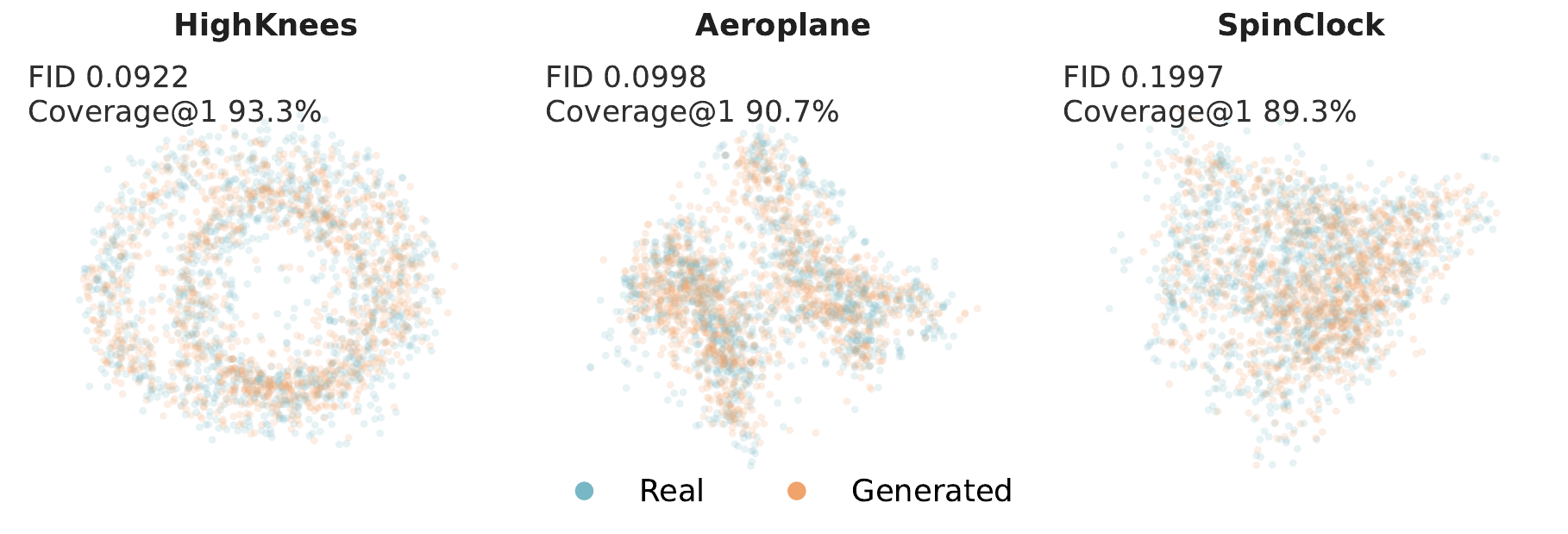}
    \caption{Distributional similarity between samples from GSI and RSI on HighKnees, Aeroplane, and SpinClock. The samples largely overlap for each style, indicating closely aligned distributions.}
\label{fig:state_init_stats}
\end{figure}

We further evaluate the quality of initial states produced by GSI compared with those from RSI. We train an autoencoder-based feature extractor and compute Frechet Inception Distance (FID)~\citep{heusel2017gans, guo2020action2motion} and Coverage at threshold 1 (Coverage@1)~\citep{li2022ganimator} in the resulting latent space. Implementation details and metric definitions are provided in \Cref{app:gsi_metrics}.
We also visualize the samples using PCA, as shown in \Cref{fig:state_init_stats}. We compute these statistics on 4096 samples of HighKnees, Aeroplane, and SpinClock, respectively, drawn from both the diffusion model and the reference dataset. The generated samples achieve FID values ranging from 0.0922 to 0.1997 and Coverage@1 above 89.3\%, indicating that the generated and reference distributions are closely aligned. The distributional similarity indicated by FID varies across styles, especially for more challenging modes such as SpinClock. Nevertheless, the experiments in \Cref{subsec:100style} show that the initial states generated by GSI is sufficient for training stylized policies, including SpinClock.

By leveraging a generative motion prior for both reward evaluation and state initialization, the original motion dataset can be completely discarded after prior construction. This enables efficient learning of natural and task-appropriate behaviors without reliance on original motion data. Such modularity facilitates flexible deployment of prior-based imitation learning frameworks and may also offer advantages in settings with data privacy constraints.

\section{Discussion and Limitations}
In this work, we presented \emph{Score-Matching Motion Priors (SMP)}, a \emph{reusable} and \emph{modular} motion prior for physics-based character animation based on score-matching. Once trained, the learned motion prior can be reused to train control policies for diverse tasks, guiding the policies towards natural behaviors that match the reference distribution. Our priors can be effectively used without the need to retain the original motion dataset. Test-time diffusion techniques, such as classifier-free guidance, can be applied to shape the base motion prior and produce novel stylistic priors that enable agents to perform tasks in specific styles. 
We demonstrate the effectiveness of our method across a diverse variety of settings, ranging from single-character behaviors to human-object interactions. SMPs can be effectively constructed from a wide spectrum of different datasets, with as few as 3 seconds of motion clips to relatively large-scale (20-hour) motion datasets.

In our experiments, we demonstrate that reinforcement learning with score distillation sampling objectives can produce motions of comparable quality to adversarial imitation learning, without the need to continuously update the prior during policy training. SMP also demonstrates higher sample efficiency than adversarial priors in most scenarios, which may be in part due to the more stable stationary reward function from SMP compared to adversarial reward models. However, as with many mode-seeking objectives, policies trained with SMP are susceptible to mode-collapse, converging to a small subset of skills rather than modeling the full distribution of possible behaviors for a particular task. This issue becomes more pronounced when training with large-scale datasets. Possible directions for increasing behavioral diversity include using generative policies or incorporating fine-grained conditioning signals into the diffusion model to explicitly specify which skills should be used by the policy. For state initialization, vanilla GSI may generate invalid motions, such as those with self-collisions, which can introduce simulation instabilities during training. Combining GSI with diffusion sampling techniques, such as guidance, may provide an effective methods to mitigate invalid states, while also serving as a promising direction for prioritized sampling and mining of difficult motions.

While our work primarily focuses on humanoid motion control, we believe SMP can also be applied to a broader range of control problems. We hope this work opens new directions toward building general motion priors that enable more versatile controllers for physics-based character animation and robotics beyond motion tracking.

\section*{Acknowledgments}
This work was supported by Sony Interactive Entertainment, NSERC (RGPIN-2015-04843), and the National Research Council Canada (AI4D-166). We would like to thank Michiel van de Panne, Amit H. Bermano, and Alejandro Escontrela for their insights and discussions.

\bibliographystyle{ACM-Reference-Format}
\bibliography{main}

\clearpage
\appendix
\section*{Appendix}
\section{Network Architecture}
\begin{figure}[t]
    \centering
    \includegraphics[width=0.9\linewidth]{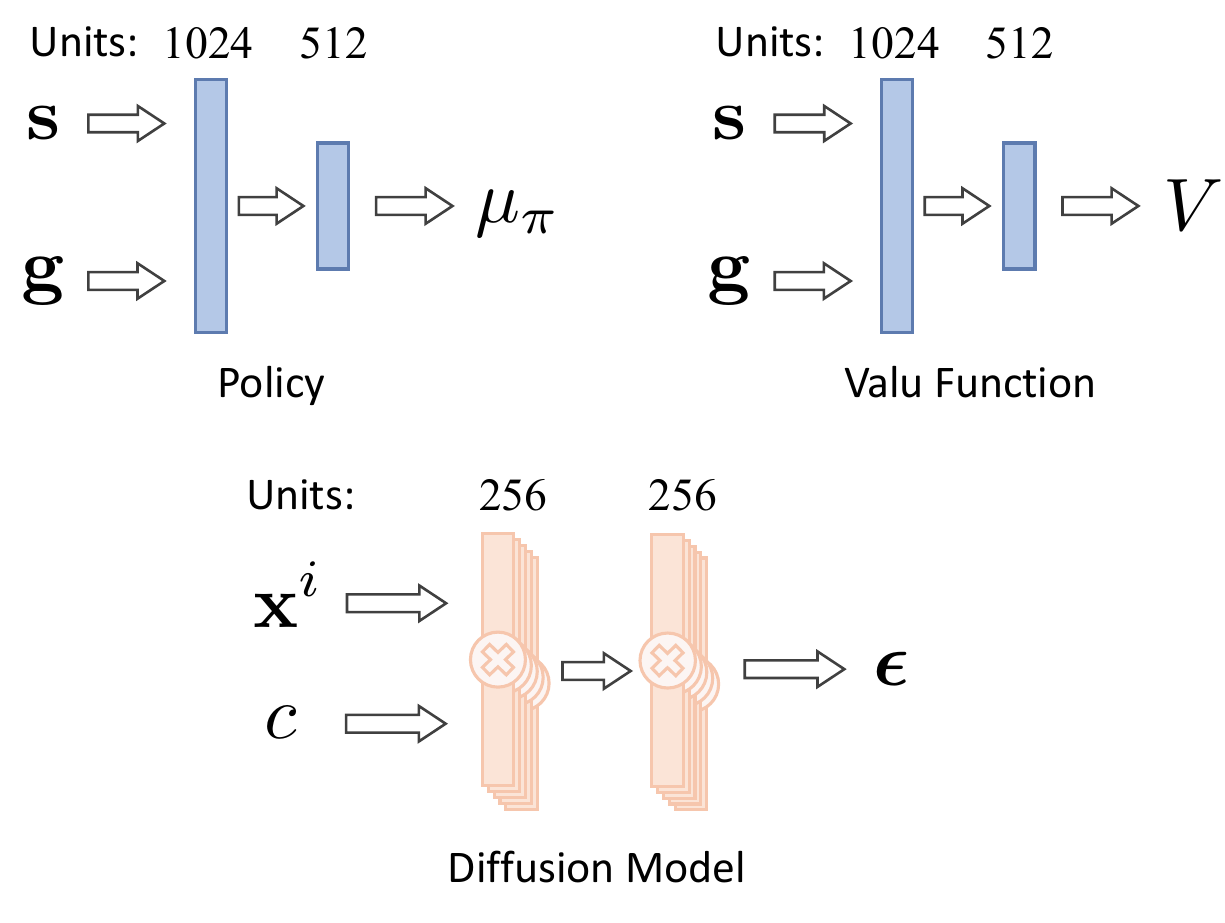}
    \vspace{-0.5cm}
    \caption{Network architectures used to model each component of the system. The policy and value functions consist of fully connected layers. The diffusion model is implemented using Transformer blocks with self-attention and adaptive normalization.}
    \label{fig:model_arch}
\end{figure}

A schematic illustration of the network architectures used for the different components of our system is shown in \Cref{fig:model_arch}.
The policy is parameterized by a neural network that maps a state $\rvs$ and task goal $g$ to a Gaussian action distribution $\pi(\rva | \rvs, g) = \mathcal{N}\!\left(\mu_\pi(\rvs, g), \Sigma_\pi\right),$
where the mean $\mu_\pi(\rvs, g)$ is input-dependent and the covariance $\Sigma_\pi$ is a fixed diagonal matrix.
The policy mean is produced by a fully connected network with two hidden layers of sizes $[1024, 512]$, followed by a linear output layer.
The value function $V(\rvs, g)$ is modeled using a similar architecture, but with a single linear output unit.
The diffusion model is implemented using two Transformer encoder blocks with multi-head self-attention, feed-forward layers, and adaptive normalization.
Each self-attention block uses 4 attention heads with a feature dimension of 64 per head, resulting in an internal embedding dimension of 256. It takes as input a noised motion clip of 10 consecutive frames, $\rvx := (\rvs_{t-8}, \dots, \rvs_{t+1})$, optionally conditioned on a style label $c$, and predicts the score $\boldsymbol{\epsilon}$.

\section{Experimental Setup}
All environments are simulated using Isaac Gym~\citep{makoviychuk2021isaac}, and all neural networks are implemented in PyTorch.
During training, 1024 environments are simulated in parallel on a single NVIDIA A100, H100, or RTX~4090 GPU.
Most policies converge within 12 hours of training.
The reinforcement learning hyperparameters are provided in \Cref{tab:suppRLParams}.
The datasets used for the object carrying and stair traversal tasks are internally collected.

\section{Single-Clip Imitation}
\label{app:single_clip}
To evaluate the performance of the various methods in imitating individual motion clips, we record the position tracking error $e^{\mathrm{pos}}_t$, which measures the differences in both the root position and relative joint positions between the simulated character and the reference motion:
\begin{align}
\label{Eq:pos_tracking_err}
e^\mathrm{pos}_t = \frac{1}{N^\mathrm{joint} + 1} \bigg( & \sum\limits_{j \in \mathrm{joints}} \left\|(\hat{\rvs}^j_t - \hat{\rvs}^\mathrm{root}_t) - (\rvs^j_t - \rvs^\mathrm{root}_t)\right\|_2^2 \nonumber \\
& + \left\|\hat{\rvs}^\mathrm{root}_t - \rvs^\mathrm{root}_t\right\|_2^2 \bigg),
\end{align}
where $\rvs^j_t$ and $\hat{\rvs}^j_t$ denote the 3D Cartesian position of joint $j$ in the simulated and reference motions, respectively. $N^\mathrm{joint}$ is the number of joints in the character.

\section{Task Implementation}
\label{app:task_implementation}
In this section, we describe the implementation details of each task, including hyperparameters and reward design.

\subsection{Target Location}
In the \textit{Target Location} task, the character is required to reach a user-specified target location $\mathbf{s}_t^{tar}$. The task reward function is formulated as,
\begin{equation}
r_t^{g\text{-}l} =
\exp\!\left(-0.5 \left\lVert \mathbf{s}_t^{tar\_2d} - \mathbf{s}_t^{root\_2d} \right\rVert^2 \right).
\end{equation}
The overall reward is then, 
\begin{equation}
r_t^{l} = 0.7 r_t^{g\text{-}l} + 0.9 r^{\text{smp}},
\end{equation}
where $r^{\text{smp}}$ is computed via \Cref{eq:averageprior}, and the SDS loss weight is set to $w^s = 4$ .

\subsection{Steering}
The \textit{Steering} task requires the character to track three commands: a desired horizontal moving direction $\mathbf{d}_t^{vel}$, a desired facing direction $\mathbf{d}_t^{face}$, and a target speed $v^{\star}_t$. To encourage the character to move in the correct direction at the desired speed while also orienting its body appropriately, we define the steering reward as a combination of a velocity tracking term and a facing alignment term:
\begin{equation}
    r^{g\text{-}s}_t = 0.5\exp\left(-2 (e^{\parallel}_t + 0.1 e^{\perp}_t)\right) +0.5\max({\mathbf{d}_t^{face}}^\top {\mathbf{h}}_t, 0).
\end{equation}
Here, ${\mathbf{h}}_t$ denotes the character’s current horizontal heading direction. We decompose the velocity tracking error into an orthogonal component,
\begin{equation}
    e^{\perp}_t = \left\lVert\dot{\mathbf{s}}_t^{root\_2d} - {\mathbf{d}_t^{vel}}^\top \dot{\mathbf{s}}_t^{root\_2d}\mathbf{d}_t^{vel}\right\lVert^2,
\end{equation}
and a parallel component,
\begin{equation}
    e^{\parallel}_t = (v^{\star}_t - {\mathbf{d}_t^{vel}}^\top \dot{\mathbf{s}}_t^{root\_2d})^2.
\end{equation}
With that, the overall reward for training policies on the \textit{Steering} task is \begin{equation}
r_t^{s} = 0.5\,r_t^{g\text{-}s} + 0.5\,r^{\text{smp}},
\end{equation}
where the scaling parameter is $w^s = 6$ in $r^{\text{smp}}$.

\subsection{Object Carry}
The \textit{Object Carry} task requires the character to walk to an object, lift it from the ground, and carry it to a user-specified target location $\rvs_t^{tar}$.
Following \citet{pan2025tokenhsi}, we design the task reward as a composition of several stage-specific components that guide the character through this multi-stage task.

The $r_t^{c\text{-}walk}$ reward guides the character to walk in the direction of $\mathbf{d}_t^{root\_obj}$, a unit horizontal vector pointing from the character's root position $\mathbf{s}_t^{root\_2d}$ to the object position $\mathbf{s}_t^{obj\_2d}$.
\begin{equation}
r_t^{c\text{-}walk} =
\begin{cases}
1.0, \left\lVert \mathbf{s}_t^{obj\_2d} - \mathbf{s}_t^{root\_2d} \right\rVert^2 < 0.5 \\[6pt]
\exp\!\left(-10 \left\lVert 0.2 - \mathbf{d}_t^{root\_obj} \cdot \dot{\mathbf{s}}_t^{root\_2d} \right\lVert \right), \text{otherwise}
\end{cases}
\end{equation}

Once the character is sufficiently close to the object, it is encouraged to pick the object up from the ground using its hands through $r_t^{c\text{-}hand}$ and $r_t^{c\text{-}pick}$. Here, $\mathbf{s}_t^{hand}$ denotes the midpoint between the character's hands, and $\mathbf{s}_t^{obj\_z}$ denotes the height of the object.
\begin{equation}
r_t^{c\text{-}hand} =
\begin{cases}
0.0, \left\lVert \mathbf{s}_t^{obj\_2d} - \mathbf{s}_t^{root\_2d} \right\rVert^2 > 0.5\\
\exp\!\left(-2 \left\lVert \mathbf{s}_t^{obj} - \mathbf{s}_t^{hand} \right\rVert^2 \right) , \text{otherwise} \\[6pt]
\end{cases}
\end{equation}

\begin{equation}
r_t^{c\text{-}pick} =
\begin{cases}
0.0, \left\lVert \mathbf{s}_t^{obj} - \mathbf{s}_t^{hand} \right\rVert^2 > 0.1\\
\exp\!\left(-5 \left\lVert1.0 - \dot{\mathbf{s}}_t^{obj\_z}\right\lVert \right),\text{otherwise}
\end{cases}
\end{equation}

Additionally, $r_t^{c\text{-}height}$ encourages the character to maintain the object above a minimum height of 0.9m, while $r_t^{c\text{-}carry}$ incentivizes transporting the object toward the target location $\rvs_t^{tar}$. 
\begin{equation}
r_t^{c\text{-}height} =
\begin{cases}
0.0, \left\lVert \mathbf{s}_t^{obj} - \mathbf{s}_t^{hand} \right\rVert^2 > 0.1\\
\exp\!\left(-7 \left\lVert0.9 - \mathbf{s}_t^{obj\_z}\right\lVert^2 \right),\text{otherwise}
\end{cases}
\end{equation}

\begin{equation}
r_t^{c\text{-}carry} =
\begin{cases}
0.2 r_t^{c\text{-}far} + 0.1 r_t^{c\text{-}near}, \left\lVert \mathbf{s}_t^{obj\_2d} - \mathbf{s}_t^{tar\_2d} \right\rVert^2 > 0.25 \\
0.2 + 0.1 r_t^{c\text{-}near}, \text{otherwise}\\[6pt] 
\end{cases}
\end{equation}
$r_t^{c-carry}$ consists of two components, $r_t^{c\text{-}far}$ and $r_t^{c\text{-}near}$. $r_t^{c\text{-}far}$ incentives the object to move along the direction of $\mathbf{d}_t^{obj\_tar}$, which is a unit horizontal vector pointing from $\mathbf{s}_t^{obj}$ to $\mathbf{s}_t^{tar}$. It plays a more important role when the object is still far away from the target.
\begin{equation}
r_t^{c\text{-}far} = \exp\!\left(-5 \left\lVert1.0 - \mathbf{d}_t^{obj\_tar} \cdot \dot{\mathbf{s}}_t^{obj\_2d} \right\lVert \right),
\end{equation}
$r_t^{c\text{-}near}$ rewards the policy based on the Euclidean distance between the object and its target.
\begin{equation}
r_t^{c\text{-}near} = \exp\!\left(-0.2 \left\lVert \mathbf{s}_t^{obj\_2d} - \mathbf{s}_t^{tar\_2d} \right\rVert^2 \right),
\end{equation}

The combined task reward for the \textit{Object Carry} task is given by
\begin{equation}
r_t^{g\text{-}c} = 0.15 r_t^{c\text{-}walk} + 0.15 r_t^{c\text{-}hand} + 0.3 r_t^{c\text{-}pick} + 0.1 r_t^{c\text{-}height} + r_t^{c\text{-}carry}.
\end{equation}

The reference motions may be imperfect in capturing precise physical interactions. These inaccuracies can be especially detrimental in contact-rich tasks like object carrying, where strict imitation of the reference motion may conflict with successful task execution. To address this, following \citet{xu2024humanvla}, we apply style reward clipping to prioritize task rewards. With style reward clipping, the overall reward for the \textit{Object Carry} task is formulated as,
\begin{equation}
r_t^c = 0.5 r_t^{g\text{-}c} + 0.5 \min(r_t^{\text{smp}}, \lambda_t),
\end{equation}
\begin{equation}
    \lambda_t = \max(r_t^{g\text{-}c}, 0.3).
\end{equation}
Here, $r_t^{\text{smp}}$ is computed according to Eq.~\ref{eq:averageprior} with $w_s = 6$.

\begin{table}[t]
    \centering
    \caption{Reinforcement learning hyperparameters used in SMP simulated character control experiments.}
    \label{tab:suppRLParams}
    \begin{tabular}{lc}
    \hline
    \textbf{Parameter} & {\bf Value} \\ 
    \hline
    $\mathcal{B}$ Experience Buffer Size  &  $1024\times 32 $ \\
    $K$ Update Minibatch Size &  $1024\times 4$  \\
    $\pi$ Policy Stepsize&  $1 \times 10^{-4}$  \\
    $V$ Value Stepsize &  $1 \times 10^{-4}$  \\
    $\gamma$ Discount&  $0.99$  \\
    SGD Momentum &  $0.9$  \\
    GAE($\lambda$) &  $0.95$  \\
    TD($\lambda$) &  $0.95$  \\
    PPO Clip Threshold &  $0.2$  \\ 
    \hline
    \end{tabular}
\end{table}

\subsection{Dodgeball}
In the \emph{Dodgeball} task, a ball is launched toward the character from a random position $8\text{-}10\,\mathrm{m}$ away, with a launch speed sampled uniformly from $[20, 25]\,\mathrm{m/s}$. The ball is relaunched at a random time within a $3$-second window. If the character is hit by the ball, the episode terminates early and the agent receives zero reward for all remaining timesteps as a penalty.

To encourage the character to avoid being hit, we define a dodge reward based on the distance between the character’s root and the ball:
\begin{equation}
r_t^{g\text{-}d}
=
1 -
\exp\!\left(
-0.3 \left(
\left\lVert
\mathbf{s}_t^{\text{root\_3d}} -
\mathbf{s}_t^{\text{ball\_3d}}
\right\rVert_2
- 1.5
\right)
\right),
\end{equation}
where the exponential form emphasizes rapid penalty growth when the ball approaches the character, while yielding diminishing penalties once a safe distance is reached.

The final reward used for training dodgeball policies is a weighted combination of the dodge reward and the SMP prior reward:
\begin{equation}
r_t^{d}
=
0.6\, r_t^{g\text{-}d}
+
1.2\, r_t^{\text{smp}},
\end{equation}
where the SDS error scaling parameter is set to $w^s = 4$ in $r^{\text{smp}}$.

\subsection{Getup}
To learn the \emph{Get-up} skill, the character is initialized from arbitrary fallen states and trained to recover to a standing pose. To improve exploration efficiency while still enabling recovery from diverse fallen conditions, we initialize the character from random frames of reference get-up motions with $90\%$ probability, and from augmented random fallen states with the remaining $10\%$ probability.

The motion prior reward $r^{\text{smp}}$ guides the learning of complex recovery skills, such as rolling over, stepping out for support, and lifting the body, which are difficult to design manually using task-specific heuristics. On the contrary, the task reward $r_t^{g\text{-}up}$ encourages functional recovery by guiding the character to raise its body toward a standing configuration.

We design the task reward as a combination of height and vertical velocity terms:
\begin{equation}
r_t^{g\text{-}up} = 0.3\, r_t^{g\text{-}uph} + 0.7\, r_t^{g\text{-}upv},
\end{equation}
where the height reward is defined as
\begin{equation}
r_t^{g\text{-}uph} =
\exp\!\left(-6\,\min(1.2 - \rvs_t^{\text{head\_h}}, 0)^2\right),
\end{equation}
and the vertical velocity reward is given by
\begin{equation}
r_t^{g\text{-}upv} =
\begin{cases}
\exp\!\left(-100\,\min(0.25 - \dot{\rvs}_t^{\text{root\_h}}, 0)^2\right),
& \rvs_t^{\text{head\_h}} < 0.6, \\[4pt]
1, & \text{otherwise}.
\end{cases}
\end{equation}
The final reward used to train the \emph{Get-up} policy is
\begin{equation}
r_t^{\text{up}}
=
0.2\, r_t^{g\text{-}up}
+
0.8\, r_t^{\text{smp}},
\end{equation}
where the SDS weight is set to $w^s = 8$ in the prior reward $r^{\text{smp}}$.

\subsection{Stair Traversal}
The \emph{Stair Traversal} task requires the character to walk up and then down a series of steps while tracking a target planar velocity.
The staircase consists of 5 steps with a step height $0.175\,\mathrm{m}$, step depth $0.3\,\mathrm{m}$, and step width $0.51\,\mathrm{m}$.
The target velocity $\mathbf{v}^{\text{tar}}$ is set to $=1.0\,\mathrm{m/s}$ along the x-axis.
The goal observation $\mathbf{g}_t=\hat{\mathbf{v}}^{\text{tar}}_t$ consists of the target planar velocity recorded in the character's local coordinate frame as a 2D vector. 

The reward function encourages the character to travel at the target velocity while facing the target direction of travel.
First, a velocity tracking reward encourages the character to move at the desired velocity:
\begin{equation}
r_t^{st\text{-}vel} =
\begin{cases}
0, & \mathbf{v}_t^{\text{root}}\cdot \mathbf{v}^{\text{tar}} \le 0,\\[4pt]
\exp\!\Bigl(-2 \left\lVert \mathbf{v}_t^{\text{root}}-\mathbf{v}^{\text{tar}} \right\rVert^2 \Bigr), & \text{otherwise},
\end{cases}
\end{equation}
where $\mathbf{v}_t^{\text{root}}$ is the character's 2D root velocity along the horizontal plane.
An additional target facing reward is included to encourage the character to face the target direction of travel,
\begin{equation}
r_t^{st\text{-}face} = \max\!\left(0,\ \mathbf{d}^{\text{tar}}\cdot \mathbf{d}_t^\text{root}\right),
\end{equation}
where $\mathbf{d}_t^\text{root}$ is the heading direction of the character's root, and $\mathbf{d}^{\text{tar}} = \mathbf{v}^{\text{tar}} / \lVert \mathbf{v}^{\text{tar}} \rVert$ denotes the target direction of travel.
The task reward is then given by a weighted combination of the two reward terms:
\begin{equation}
r_t^{g\text{-}st} = 0.7\, r_t^{st\text{-}vel} + 0.3\, r_t^{st\text{-}face}.
\end{equation}
The task reward is then combined with the SMP reward via:
\begin{equation}
r_t^{st} = 0.5\, r_t^{g\text{-}st} + 0.5\, r_t^{\text{smp}},
\end{equation}
where $r_t^{\text{smp}}$ is computed from the SDS loss with a scale parameter $w^{s}=8$.

\section{Style Composition}
We demonstrate a practical approach for creating new motion styles from a base SMP by composing diffusion model predictions from different styles for the upper and lower body:
\begin{equation*}
f_{\mathrm{comp}}
= M_{\mathrm{upper}} \odot f(\rvx^i, c_{\mathrm{style1}})
+ M_{\mathrm{lower}} \odot f(\rvx^i, c_{\mathrm{style2}}),
\end{equation*}
where $M_{\mathrm{upper}}$ and $M_{\mathrm{lower}}$ are binary masks that select upper- and lower-body motion features, respectively.
Although these predictions are computed from the same noised sample $\rvx^i$ and combined in the $\boldsymbol{\epsilon}$-space, composing styles that are highly dissimilar or contradictory can be challenging. In practice, directly blending upper- and lower-body priors may lead to diluted styles, as the resulting composite score may not correspond to a smoothly realizable motion.

To address this issue, we introduce \emph{multi-step score matching} (MSM), where the score $\hat{\boldsymbol{\epsilon}}$ is estimated through a multi-step reverse diffusion process rather than a single-step prediction. We employ a DDIM sampler with a partial denoising schedule, using a stride of two over three reverse steps~\citep{song2020denoising}. MSM offers two key advantages. First, the iterative reverse diffusion process naturally blends the composed scores over multiple denoising steps, resulting in more coherent composite styles. Second, the resulting multi-step score estimate is more accurate than a single-step prediction~\citep{liang2024luciddreamer}.

This improved expressiveness comes at the cost of increased computation during reward evaluation. For example, training a policy to perform the composite ``\emph{AeroPlane + HighKnees}'' style using MSM requires approximately 16 hours of training on a single RTX 4090 GPU. All such operations are applied exclusively during reward computation and do not modify the policy architecture or action space. Once training is complete, the policy alone can complete tasks while exhibiting the learned natural and expressive behavioral style.

\section{Real-World Robotic Deployment}
\begin{table}[t]
    \centering
    \caption{Domain randomization hyperparameters and regularization terms used in SMP real-world robot deployment experiments.}
    \label{tab:suppDomainRandParams}
    \begin{tabular}{lc}
        \hline
        \textbf{Term} & {\bf Value} \\ \hline
        \multicolumn{2}{c}{\textbf{Observation Noise}} \\
        \hline
        Joint Velocity Noise &  $\mathcal{U}(-0.5, 0.5)$ rad/s \\
        Joint Position Noise &  $\mathcal{U}(-0.025, 0.025)$ rad \\
        Root Angular Velocity Noise &  $\mathcal{U}(-0.25, 0.25)$ rad/s \\
        Root Rotation Noise &  $\mathcal{U}(-5, 5)$ deg \\\hline
        \multicolumn{2}{c}{\textbf{Physical Properties}} \\
        \hline
        Friction &  $\mathcal{U}(0.1, 2)$ \\
        COM Offset &  $\mathcal{U}(-0.1, 0.1)$ m \\
        Kp, Kd scale & $\mathcal{U}(0.8, 1.25)$ \\
        Push Robot & $\mathcal{U}(-0.5, 0.5)$ m/s \\
        Link Mass &  $\mathcal{U}(0.9, 1.1)$  \\\hline
        \multicolumn{2}{c}{\textbf{Regularization}} \\
        \hline
        Action Rate &  $-0.04$  \\
        Self Contact &  $-0.2$  \\
        \hline
    \end{tabular}
\end{table}

To facilitate real-world deployment, we apply domain randomization techniques during training in simulation. This includes applying external perturbations, injecting noise into the policy’s observations, and randomizing physical parameters of the simulator.
Table~\ref{tab:suppDomainRandParams} provides a detailed summary of the domain randomization hyperparameters, as well as the regularization terms and their corresponding weights used in the experiments.

\section{Quantitative Comparison of GSI and RSI}
\label{app:gsi_metrics}

To quantitatively evaluate the quality of motions used for state initialization under GSI, we compute the Frechet Inception Distance (FID)~\citep{heusel2017gans, guo2020action2motion} and Coverage at threshold 1 (Coverage@1)~\citep{li2022ganimator}, following standard protocols for evaluating kinematic motion quality.

\paragraph{Feature Extractor.}
For quantitative evaluation, we use the encoder of a trained VAE as the feature extractor. The encoder is implemented as a 4-layer DiT with 4 attention heads, which processes a normalized 10-step motion window and predicts the mean and log-variance of a 16-dimensional latent Gaussian. A lightweight MLP decoder is used to reconstruct the input motion. The VAE is trained with the standard reconstruction-plus-KL objective:
\begin{equation}
\mathcal{L}_{\mathrm{VAE}}
=
\|\hat{\rvx}-\rvx\|_2^2
+
\lambda_{\mathrm{KL}}\,
\mathrm{KL}\!\left(
q_\phi(\rvz \mid \rvx)
\,\|\, 
\mathcal{N}(\mathbf{0}, \mathbf{I})
\right),
\end{equation}
where $\lambda_{\mathrm{KL}} = 0.1$ in our implementation, and $q_\phi$ denotes the posterior distribution parameterized by the encoder with parameters $\phi$.

\paragraph{FID and Coverage.}
We sample 4096 motions from both the diffusion model and the reference dataset. Real and generated motions are normalized using the same VAE normalizer and mapped into latent features using the posterior mean:
\begin{equation}
\rvz = \mu_\phi(\tilde{\rvx}) \in \mathbb{R}^{16}.
\end{equation}

Given real features $\{\rvz_i^{r}\}_{i=1}^{N_r}$ and generated features $\{\rvz_j^{g}\}_{j=1}^{N_g}$, with empirical means and covariances $(\mu_r, \mathbf{\Sigma}_r)$ and $(\mu_g, \mathbf{\Sigma}_g)$, respectively, we compute FID as
\begin{equation}
\mathrm{FID}
=
\|\mu_r-\mu_g\|_2^2
+
\mathrm{Tr}\!\left(
\mathbf{\Sigma}_r+\mathbf{\Sigma}_g
-
2(\mathbf{\Sigma}_r \mathbf{\Sigma}_g)^{1/2}
\right).
\end{equation}

For coverage, we first compute the nearest-neighbor distance from each real feature to the generated set in the latent space:
\begin{equation}
d_i = \min_{1 \le j \le N_g} \|\rvz_i^{r} - \rvz_j^{g}\|_2,
\end{equation}
and then define coverage at threshold $\tau$ as
\begin{equation}
\mathrm{Coverage}@\tau
=
\frac{100}{N_r}
\sum_{i=1}^{N_r}
\mathds{1}[d_i \le \tau].
\end{equation}
We report Coverage@1 with $\tau = 1.0$, following \citet{li2022ganimator, raab2024single}.

\end{document}